\documentstyle[psfig]{mn}
\psfigurepath{plots/}

\def\snr{G320.4--01.2}
\def\msh{MSH~15--5{\em 2}}
\def\rcw{RCW~89}
\def\psr{B1509--58}

\newcommand\HI{H\,{\sc i}}

\newcommand\kms{km~s$^{-1}$}
\newcommand\etal{et~al. }
\newcommand{\miriad}{{\tt MIRIAD}}

\begin{document}
\label{firstpage}
\title[Radio observations of SNR~\snr\ and PSR~\psr]
{SNR~\snr\ and PSR~\psr: New radio observations of a complex interacting
system}
\author[B. Gaensler \etal]
{B. M. Gaensler,$^{1,2}$\thanks{Current address:
Center for Space Research, Massachusetts Institute of Technology,
Cambridge, MA 02139, USA. Email: bmg@space.mit.edu}
K. T. S. Brazier,$^3$ R. N. Manchester,$^2$ S. Johnston$^4$
\newauthor and A. J. Green$^1$ \\
$^1$Astrophysics Department, School of
Physics A29, University of Sydney, NSW 2006, Australia \\
$^2$Australia Telescope National Facility, CSIRO, PO Box 76,
Epping, NSW 1710, Australia \\
$^3$Department of Physics, University of Durham, South Road, Durham
DH1~3LE, United Kingdom \\
$^4$Research Centre for Theoretical Astrophysics, 
University of Sydney, NSW 2006, Australia}

\pagerange{\pageref{firstpage}--\pageref{lastpage}}
\pubyear{1998}

\maketitle
\begin{abstract}

We describe radio continuum and spectral line observations 
of the source \snr\ (\msh) and the coincident young pulsar \psr\ made
with the Australia Telescope Compact Array.
Based on a comparison between X-ray and radio observations,
we argue that the two main radio components of
\snr\ are a
single supernova remnant (SNR), 
which \HI\ absorption indicates is at a distance of
$5.2\pm1.4$~kpc.  A high-resolution correspondence between radio and
X-rays argues that the pulsar is interacting with the SNR via an
opposed pair of collimated outflows.  The outflow itself is seen as an
elongated X-ray feature surrounded by a highly polarized radio sheath, while
the interaction with the SNR manifests itself as a ring of radio/X-ray
knots within the optical nebula \rcw. We reject the hypothesis that the
pulsar outflow powers the entire \rcw\ region.

SNR~\snr\ and PSR~\psr\ agree in distance and in rotation measure,
and appear to be interacting. We thus conclude that the two objects
are associated and have an age of $\la$1700~yr.  
We propose that the SNR resulted from
a high-energy or low-mass supernova which occurred near the edge of
an elongated cavity.  Such a model can account for the SNR's bilateral
appearance, its large apparent age, the significant offset of the pulsar
from the SNR's centre and the faintness of the pulsar-powered
nebula at radio wavelengths.

\end{abstract}


\section{Introduction}
\label{sec_g320_intro}

A massive star ends its life in a supernova. This 
produces a supernova remnant (SNR) and often also
an associated neutron star,
the latter
sometimes observable as a pulsar. However, associations between SNRs
and pulsars are rare: 30 years of effort have provided fewer than
ten convincing cases (e.g. Kaspi 1996\nocite{kas96}).  
Such associations can clarify questions
regarding pulsar velocities, ages, magnetic fields and initial spin
periods, and can help us to understand the evolution and appearance
of SNRs. Thus establishing the validity of an association is of great
interest, as are subsequent studies of a particular pulsar/remnant
pairing.

Of such systems, one of the best studied but least understood is that
involving the peculiar
SNR~\snr\ and the energetic young pulsar \psr. In this paper we
present a new radio study of this intriguing pair of objects.

\subsection{Previous Observations}

\subsubsection{Radio}

\begin{figure}
\centerline{\psfig{file=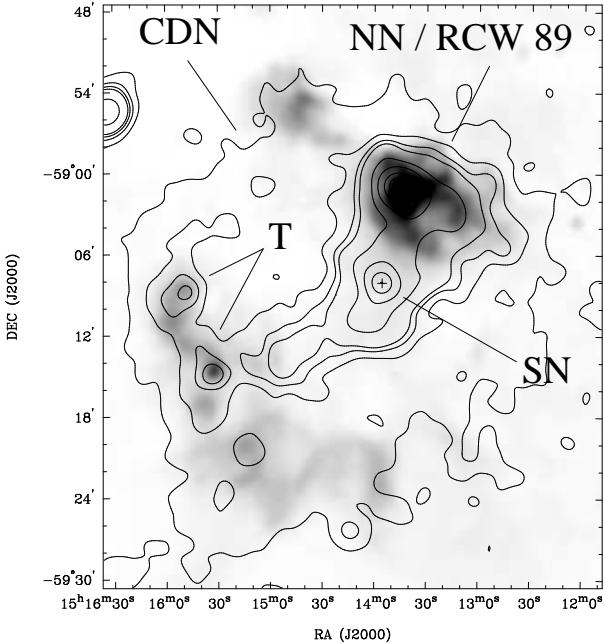,width=8cm,angle=270}}
\caption{A radio/X-ray comparison of \protect\snr. 
The grey-scale corresponds to 36-cm MOST observations 
\protect\cite{wg96},
while the contours represent 
smoothed {\em ROSAT} Position Sensitive Proportional Counter (PSPC) 
data of Trussoni \etal \protect\shortcite{tmc+96}
[their Figure~2(a)]. Contour levels (in arbitrary units) are
at levels of 0.5, 1, 1.5, 2, 5, 10, 20, 30 and 40. 
Selected regions designated by Trussoni \etal
\protect\shortcite{tmc+96} are indicated. Here (and in
subsequent figures), the position of
PSR~\psr\ is marked by a ``+'' symbol.
G320.6--01.6 is to the south-east of \snr\ and
is outside the field of view shown here.}
\label{fig_g320_pspc}
\end{figure}

The radio source~\snr\ (\msh, Kes~23) was detected in the earliest radio
surveys of the southern sky \cite{msh60,kom66,hil68,kes68,dtg69}, and
was soon classified as a SNR on the basis of its non-thermal
spectrum and lack of hydrogen recombination lines
\cite{mwgm69,mil70,sg70c}.  Early images of the source
\cite{sg70b,mil72b,md75} showed (at least) two distinct components: a
bright double source to the north-west, and fainter extended structure
to the south-east.  Subsequent observations at sub-arcmin
resolution \cite{cmw81,md83,wg96} showed the north-western component to
be dominated by a central clumpy ring, and the south-eastern
source to be comprised of complex filamentary structure, as shown
in Figure~\ref{fig_g320_pspc}. Both
components are extended along a direction parallel to the Galactic Plane,
with a gap between them. 

The radio morphology of \snr\ is unusual. Its two well-separated
components can only be considered a single shell in the broadest of
interpretations; the unusual collections of knots and filaments, and
the distinct gaps in emission to the north-east and south-west, need to
be explained. The region is probably the site of many supernovae in the
recent past \cite{lgg87}, and one possibility is that the components of
\snr\ may actually be multiple SNRs.
Wide-field observations of the region
\cite{mch93,wg96} reveal a pair of polarized radio arcs $\sim$30~arcmin
to the south-east of \snr, which have been designated 
G320.6--01.6.  It is not clear whether this 
structure is another part of \snr\ or is a separate SNR.

\subsubsection{X-ray}

The first X-ray image of the region revealed a surprise:  a 150-ms pulsar
\psr\ within the SNR \cite{sh82}. The pulsar was
subsequently detected in the
radio \cite{mtd82} and $\gamma$-ray \cite{wff+92,umw+93}
bands.
Radio timing observations
\cite{mdn85,kms+94} show PSR~\psr\ to have the highest period derivative
($\dot{P} = 1.5 \times 10^{-12}$), the second lowest characteristic
age ($\tau_c = 1700$~yr) and the third highest spin-down luminosity
($\dot{E} = 1.8 \times 10^{37}$~erg~s$^{-1}$) of any known pulsar.

The X-ray morphology of the SNR is shown in Figure~\ref{fig_g320_pspc}.
The pulsar is embedded in a 5--10 arcmin
non-thermal nebula (region ``SN'' of Figure~\ref{fig_g320_pspc})
\cite{shss84,gcm+95,mbg+97}, presumed to be
synchrotron emission from a pulsar wind nebula (PWN).  
There is no obvious radio counterpart to this PWN.  
To the north-west of the pulsar
is an extended thermal X-ray source (region ``NN'') 
\cite{shmc83,tmc+96,tkyb96}, whose
position and morphology closely correspond to the brightest radio
emission \cite{bb97}.  A diffuse bridge of emission joins the ``SN''
and ``NN'' regions. Another bridge connects the ``SN'' region
to X-ray emission south-east of the
pulsar (region ``T''). Region ``T'' appears to coincide in position and
morphology with the south-eastern component of the
radio SNR \cite{tmc+96}. The lowest level X-ray contour 
envelopes the entire SNR (region ``CDN'').

\subsubsection{Optical and Infrared}

The bright north-western X-ray and radio source coincides with the collection
of irregular filaments which make up the
H$\alpha$ nebula \rcw\ \cite{rcw60,shmc83}. Other nearby nebulosities
are probably unrelated to the SNR \cite{lgg87}. The infrared point source
IRAS~15099--5856 is coincident with the pulsar, but its nature is
uncertain \cite{are91}.

\subsection{Are \snr\ and \psr\ associated?}
\label{sec_g320_intro_questions_assoc}

PSR~\psr\ is a young pulsar located near the centre of a SNR.  At first
glance, a physical association with \snr\ seems assured.  However, such
a claim should be treated with caution: a young pulsar need not have
an associated SNR \cite{ksbg80,bgl89}, and line-of-sight coincidences
are rife (Gaensler \& Johnston 1995a,b\nocite{gj95b,gj95c}).  There are
several criteria by which an association can be judged. The most useful
are agreement in distance and age estimates, 
a reasonable implied transverse velocity for the pulsar, 
evidence from the pulsar's proper motion, 
and any indication of an interaction between the pulsar and the SNR.

\HI\ absorption towards \rcw\ \cite{cmr+75} puts the SNR at a distance
of 4.2~kpc, slightly less than the pulsar's distance
of 5.9$\pm$0.6~kpc as derived from its dispersion measure \cite{tc93}.
A more serious discrepancy lies in the ages of the two systems: unless
one invokes a time-variable braking index, PSR~\psr\ can be no older
than 1700~yr \cite{kms+94}. But assuming standard supernova and interstellar
medium (ISM) parameters,
Seward \etal \shortcite{shmc83} calculate an age for the SNR in the
range $(6-21) \times 10^3$~yr. Proper motion measurements of the optical
filaments in \rcw\ also imply a remnant age much greater than that of
the pulsar \cite{vk84}. 
One way of reconciling the SNR and pulsar ages
is by invoking rapid expansion of the SNR shell into a cavity, followed
by an encounter with dense surrounding material \cite{shmc83,man92b}.

Calculating the transverse pulsar velocity implied by an association
requires a determination of the pulsar birthplace.  This is
straightforward when the SNR is a near-circular shell, but is more
difficult for the complex morphology of \snr. Suffice to say that the
pulsar's current position well within the SNR's perimeter implies
a projected velocity less than $\sim$3000~\kms.
No proper motion has yet been detected \cite{kms+94}.

In order to explain \rcw's bright appearance in X-rays, Seward \etal
\shortcite{shmc83} suggested that it is powered by an outflow of energy
from the pulsar. Tamura \etal \shortcite{tkyb96} revive this
interpretation, showing that the X-ray bridge joining the pulsar and
\rcw\ has a non-thermal spectrum. They propose that this bridge is a
jet flowing out from the pulsar and illuminating \rcw.  Manchester \&
Durdin \shortcite{md83} and Manchester \shortcite{man87}
describe a related idea, noting that the two
components of the radio remnant can be mapped onto opposed cones
emanating from PSR~\psr. They suggest that the radio morphology of the
SNR may be generated by opposed beams of particles originating from the
pulsar and colliding with the surface of a surrounding cavity.  Brazier
\& Becker \shortcite{bb97} provide X-ray evidence for this model,
pointing out that the PWN has a cross-shaped appearance, possibly
corresponding to a highly inclined version of the ``torus + jets''
geometry of the Crab Nebula in X-rays
\cite{hss+95}.  The ``jets'' align with
the axis of proposed outflow (presumed also to be the pulsar's rotation
axis) and one of them maps onto the ring of X-ray knots seen at the
peak of \rcw.

There are two main problems with the argument that the \rcw\ region  is
powered purely by an outflow from the pulsar. First, the total thermal
energy in \rcw\ ($4 \times 10^{49}$~erg; Tamura \etal
1996\nocite{tkyb96}) exceeds the {\em entire} energy lost by the pulsar
over its lifetime for any but very rapid initial periods. Second, the
mechanism by which a large thermal nebula might be created at the
termination of such an outflow is unclear \cite{tkyb96}. Both these
issues must be resolved if this model is to be considered a
possibility.

\section{New Observations}
\label{sec_g320_obs}

Clearly, there are many issues still to be resolved involving
SNR~\snr\ and PSR~\psr. This has prompted us to carry out a set of
new radio observations, using
the Australia Telescope Compact
Array (ATCA; Frater, Brooks \& Whiteoak 1992\nocite{fbw92}), a synthesis
telescope located near Narrabri, NSW, Australia.  The ATCA consists of five
moveable 22-m antennas on a 3-km east-west track, with a fixed sixth
antenna stationed 3~km further west.  
Five types of observations were carried out, as detailed in
Table~\ref{tab_g320_observations}:

\begin{enumerate}
\item continuum observations at 20~cm (1.3~GHz) and 6~cm
(4.8 and 5.8~GHz);
\item pulsar-gated observations;
\item observations in the \HI\ spectral line;
\item observations in the 1720-MHz OH maser line;
\item observations in the H140$\alpha$ recombination line (2371.1~MHz).
\end{enumerate}

\HI\ and ungated continuum observations at 20~cm consisted of two
fields covering the entire SNR.
Continuum observations at 6~cm were a mosaic of eight fields,
covering \rcw, a region around \psr\ and part of the south-eastern
component. OH, H140$\alpha$ and pulsar-gated observations consisted
of a single pointing towards \rcw. All four Stokes parameters
were recorded in continuum, while total-intensity alone was recorded
for line observations. The flux density scale of all data was tied to 
the revised scale of Reynolds \shortcite{rey94}
using observations of PKS~B1934--638.
Antenna gains were determined using regular observations of 
MRC~1613--586 (20-cm continuum, \HI\ and H140$\alpha$), 
PKS~B1740--517 (6-cm continuum) and PKS B0823--500 (OH). 

\begin{table*}
\begin{minipage}{150mm}
\caption{ATCA observations of \snr. Except where noted, all
observations were in continuum mode, involving 32 channels across
a 128 MHz bandwidth.}
\label{tab_g320_observations}
\begin{tabular}{cccccc} 
Date    & Array         & Maximum      & $\nu_1$ & 
$\nu_2$ &Time on \\
	& Config & Baseline (m) & (MHz) & (MHz) &
	Source (h) \\ \hline
1995 Nov 03 & 6A    & 5939 & 1344 & 1420$^a$ & 4 \\
1995 Nov 11 & 6A    & $"$ & 1721$^b$ & --  & 3 \\
1996 Feb 01 & 0.75B & 765 & 1344 & 1420$^a$ & 3 \\
1996 May 10 & 1.5D  & 1439 & 1376 & 1420$^a$ & 16 \\
1996 May 12 & 1.5D  & $"$ & 4790 & 5824 & 6 \\
1996 May 13 & 0.75D & 719 & 4790 & 5824 & 9 \\
1996 May 16 & 0.75D & $"$ & 1344 & 1420$^a$ & 12 \\
1996 May 17 & 0.75D & $"$ & 4790 & 5824 & 12 \\
1996 Nov 10 & 0.75A & 735 & 2371$^c$ & 1376$^d$ &  13 \\
1997 Apr 18 & 0.375 & 459 & 1344 & 1420$^a$ & 11 \\
1997 Apr 20 & 0.375 & $"$ & 4800 & 5824 & 13 \\ \hline
\end{tabular}

$^a$ \HI\ observations (1024 channels across 4 MHz) \\
$^b$ OH observations (1024 channels across 4 MHz);
these observations also formed part of the survey of Green
\etal (1997\nocite{gfgo97}), but have been reprocessed here \\
$^c$ H140$\alpha$ observations (256 channels across 8 MHz) \\
$^d$ Pulsar-gated observations (similar to continuum mode, but where
data were sampled every 4.7~ms and then folded at the apparent period
of \mbox{PSR~\psr}) \\

\end{minipage}
\end{table*}

\section{Data reduction}
\label{sec_g320_reduction}

Reduction and analysis were carried out within the \miriad\
package \cite{stw95}. Data were
edited and calibrated using standard techniques \cite{sk97}. The field
of the calibrator MRC~1613--586 contains several weak confusing
sources, and antenna gains for this source were determined using
a model consisting of both the calibrator and these weaker sources
(cf.\ Gaensler, Manchester \& Green 1998a\nocite{gmg98}).

\subsection{Continuum data}
\label{sec_g320_reduction_cont}

At both 20 and 6~cm, a mosaic image of the field was formed
using multi-frequency synthesis \cite{sw94} and uniform weighting.
Data from the 6-km antenna were excluded.  Each image was deconvolved
using the maximum entropy algorithm \cite{gd78}, all pointings in the
mosaic being handled simultaenously \cite{ssb96}.
The resulting models were
smoothed using a Gaussian restoring beam, and
then corrected for the mean primary response of the ATCA antennas and
for the appropriate mosaic pattern. The resolution and noise
in the final images are given in Table~\ref{tab_g320_snr}.

The pulsar-gated data were analysed by de-dispersing at a dispersion
measure of 253.2~pc~cm$^{-3}$ \cite{kms+94}. On- and
off-pulse datasets were formed, and the off-pulse data subtracted from 
the on-pulse data so that only pulsed emission remained.

Images of the region were also formed in Stokes $Q$, $U$ and $V$.  At
20~cm, Faraday rotation across the observing band is significant
($\sim$4 radians across 128~MHz in some regions), and can cause
depolarization of the emission.  This effect was minimised by making
multiple pairs of $Q$ and $U$ images across the 20-cm band, each of
bandwidth 8~MHz. At 6-cm bandwidth depolarization
is minimal ($<0.5$~radians), and
we made a single pair of $Q$ and $U$ images for the entire frequency
band.

Although mosaic images can be formed in polarization as described
for total-intensity data above, $Q$, $U$ and $V$ data are not positive
everywhere and thus cannot be deconvolved using maximum entropy techniques
as used for Stokes~$I$. For our 20-cm data we thus
deconvolved images of polarized emission using
the ``individual'' approach (see Sault \& Killeen 1998\nocite{sk97}),
whereby each pointing is deconvolved separately using the {\sc CLEAN}
algorithm \cite{cla80}, and final images in $Q$, $U$ or $V$ are then
formed by combining each CLEANed pointing.  However the ``individual''
approach results in an image greatly lacking in extended structure when
compared to ``joint'' deconvolution techniques \cite{cor88}.  At 20~cm
most of the emission is well sampled by the unmosaiced $u-v$ coverage,
and this effect is minimal.  However at 6~cm the image quality resulting
from the individual approach is poor. Thus at 6~cm we
formed dirty polarization images  using the joint
algorithm as in total intensity,
but made no attempt to deconvolve them.

A linear polarization image $L$ was formed from each pair of $Q$ and $U$
images and corrected for non-Gaussian noise statistics.
At 20~cm, each $L$ image was clipped  where the
polarized emission or the total intensity was less than 5$\sigma$, and
the final image formed by taking a mean across the observing band.

\subsection{Spectral Index}
\label{sec_g320_reduction_spec}

Existing spectral index studies of \snr\ have had low
spatial resolution and have produced a wide range
of conflicting
results (see du Plessis \etal 1995\nocite{ddb+95}). 
But using our high-resolution 20- and 6-cm observations, 
we can accurately calculate the
spatial distribution of $\alpha$ (where $S_\nu
\propto \nu^{\alpha}$) for \snr.  

When computing spectral indices, it is crucial that
the images being
compared contain the same spatial scales. 
The $u-v$ coverage of our two datasets is quite different,
and spatial filtering must be applied to each 
so that their spatial scales match. Assuming that
the deconvolution process
at each frequency accurately fills in the $u-v$ plane in some range
$w_{\rm min} \le (u^2 + v^2)^{1/2} \le w_{\rm max}$, this
filtering can be simply achieved by smoothing the higher resolution
image to the resolution of the lower, and discarding extended
structure in whichever image extends to shorter projected baselines.
The spatial distribution of spectral index
can then be determined using the method
of ``T--T'' (temperature-temperature)
plots \cite{tpkp62}. In this method, two images
are plotted against
each other on a pixel-by-pixel basis in some small sub-region of the
image. Provided any DC offset is constant over this region, the data
will follow a linear relationship , whose slope $m$
is related to the spectral index, $\alpha$, by

\begin{equation}
\alpha = \frac{\ln m}{\ln (\nu_1/\nu_2)}.
\label{eqn_specindex3}
\end{equation}
A linear fit to the data gives a value of $m$ and thus of $\alpha$.

We applied this method to our 20- and 6-cm data on \snr\ to produce
a pair of images of resolution 24 arcsec.
Because there are many pixels per beam area in each image, adjacent
pixels are correlated, which will cause the uncertainty in
the linear fit to a T--T plot to be underestimated
\cite{gre90}. To correct for this effect, each image was re-binned
by choosing every $i$th pixel, where there are $i$ pixels per beam in
each dimension. T--T plots of $\nu_1$ vs $\nu_2$ were then produced in
square regions of side 2 arcmin, and a linear least squares fit applied
to each plot to determine $m$, and hence $\alpha$.

\subsection{Spectral line data}

For the \HI\ observations,
continuum emission was subtracted from the line data in the $u-v$ plane
\cite{vc90,sau94}, and spectra were
then smoothed to a velocity resolution of 3.3~\kms.  Diffuse
emission was removed by discarding $u-v$ spacings shorter
than 1~k$\lambda$, and a line cube was then formed for velocities
between --200 and +200 \kms\ (LSR). The peak emission in the resulting
images was deemed sufficiently faint that deconvolution was not required.

The \HI\ cube was then weighted by multiplying it by the 20-cm continuum image.
\HI\ spectra towards sources of interest were generated by integrating
over the relevant region of the cube, and then renormalised 
to give units of fractional absorption.

The rms noise, $\sigma$, in each spectrum was determined from the flux
in line-free channels. The presence of extended \HI\ emission in the
Galactic Plane raises the thermal noise in the \HI\ line to
approximately double this value (e.g.\ Dickey 1997\nocite{dic97}), and
so we choose 6$\sigma$ as the threshold above which features are
considered believable.

Continuum emission was subtracted from OH and H140$\alpha$ data as for
\HI. Cubes were formed at various  velocity resolutions and then
searched for emission.

\section{Results}
\label{sec_g320_results}

\subsection{Continuum Images}
\label{sec_g320_results_cont}

\begin{figure*}
\begin{minipage}{160mm}
\centerline{\psfig{file=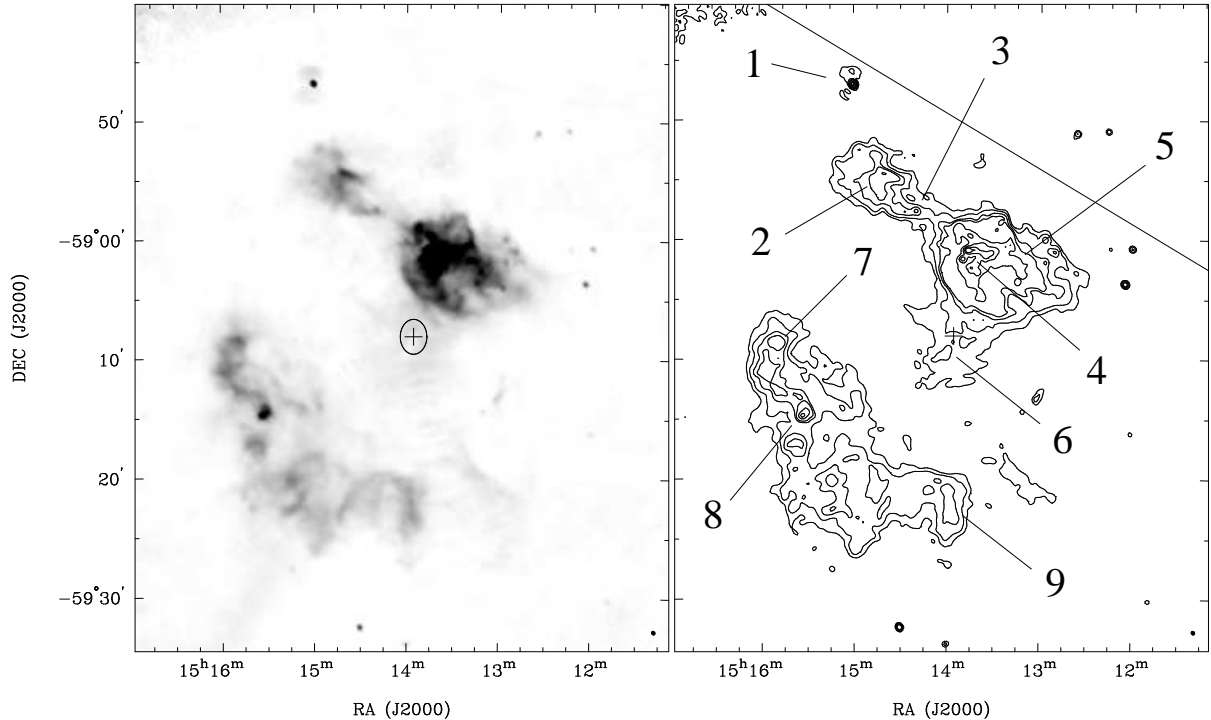,width=16cm}}
\caption{Grey-scale and contour images of SNR~\snr\ at 20~cm, corrected
for the primary beam response of the telescope.
The grey-scale range is 0 to 75~mJy~beam$^{-1}$, while the
contours are at 5, 10, 20, 30, 60, 90, 120 and 150~mJy~beam$^{-1}$. 
The FWHM of the Gaussian restoring beam is
shown at the lower right of each image. The diagonal line above the SNR
runs parallel to the Galactic Plane. Various regions of interest are marked.
The ellipse in the left panel
shows the 1$\sigma$ error ellipse for the pulsar's birth-place.}
\label{fig_g320_20cm}
\end{minipage}
\end{figure*}

\begin{figure}
\centerline{\psfig{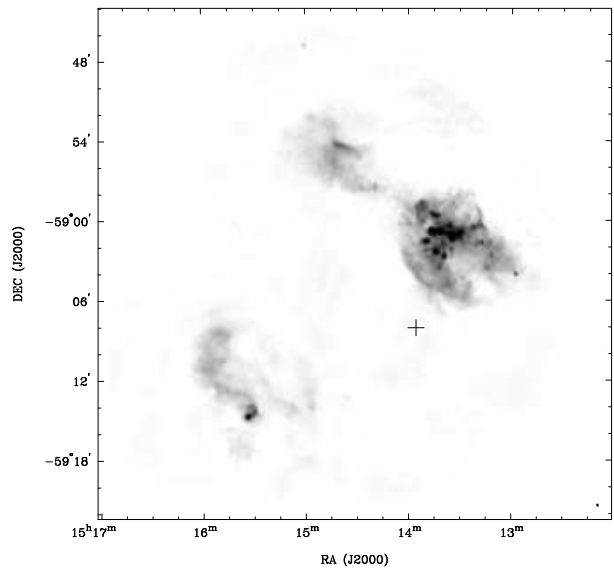}}
\caption{Grey-scale image of part of SNR~\snr\ at 6~cm. The
image has been tapered
towards the edges to give uniform noise across the image,
and shows emission in the range 0 to 20~mJy~beam$^{-1}$.
The FWHM of the Gaussian restoring beam is
shown at lower right.}
\label{fig_g320_6cm}
\end{figure}

Total-intensity images of SNR~\snr\ at 20 and 6~cm are shown in 
Figures~\ref{fig_g320_20cm}
and~\ref{fig_g320_6cm} respectively.  
The brightest region is that coincident with \rcw, and
can be seen to form a ring
of knots (region 4 in Figure~\ref{fig_g320_20cm})
superimposed on a horseshoe-shaped plateau of emission (region 5). 
A bridge of emission running parallel to the Galactic
Plane connects \rcw\ with another clump of emission, comprised of a
compact source at its south-west end (region 3) and becoming more
diffuse to the north-east (region 2).  South-east of \rcw\ is a
twisted, filamentary arc (incorporating regions 7 and 9), with one distinctly
compact region near its northern end (region~8).

The 20-cm flux density of \snr\ and its associated uncertainty are
given in Table~\ref{tab_g320_snr}, 
and were measured by integrating polygons
around each component of the source then applying a background
correction. The shortest antenna spacing at 20~cm is 31~m,
corresponding to a spatial scale of $\sim$25~arcmin. However, the mosaic
allows us to recover information on larger scales (Cornwell 1988;
Sault et al. 1996\nocite{cor88,ssb96}),
in this case up to $\sim$35~arcmin.  If the SNR contains structure
larger than this, these missing spacings will result in a significant
underestimate in the true flux density.  However, the 20-cm flux
densities in Table~\ref{tab_g320_snr}  
are in excellent agreement with single-dish
observations at this frequency \cite{mwgm69,mil72b}.  Furthermore, 
there is no negative bowl observable around the SNR, as
would be expected if large-scale structure were missing. Finally, 
our ATCA image, when smoothed to the appropriate
resolution, is indistinguishable from a 20-cm image made with the
Parkes 64-m radio telescope (A.~R.~Duncan 1996, private
communication).  All these points are good evidence that the 20~cm ATCA
observations contain all the flux and structure from the SNR.

At 6~cm, we are sensitive only to scales smaller than 9~arcmin,
even after mosaicing. 
Thus we expect the flux density
determined in Table~\ref{tab_g320_snr} (for the northern component
only) to be an underestimate. Indeed if we take the flux density
at 20~cm and scale it by a mean spectral index
$\alpha = -0.45$ (Milne \etal
1993\nocite{mch93}), we expect a 6-cm flux density
for the northern region of 15~Jy, substantially greater than
the value in Table~\ref{tab_g320_snr}. This missing flux is clearly
evident in Figure~\ref{fig_g320_6cm}, which,
compared to Figure~\ref{fig_g320_20cm},
lacks significant extended structure.

The position and 20-cm flux density of PSR~\psr\ are
given in Table~\ref{tab_g320_snr} and were
determined by fitting in the $u-v$ plane to the pulsed dataset formed
from the pulsar-gated data.  This position is marked
by a ``+'' symbol on appropriate Figures. The pulse-profile 
is similar
to that obtained by Manchester \etal \shortcite{mtd82}, consisting of a
single main pulse occupying $\sim$15 per cent (FWHM) of the period.  
The pulsar was not detected in (ungated) 6-cm observations.

The position of the pulsar given in Table~\ref{tab_g320_snr} 
is consistent, within the
uncertainties, with that derived by Manchester \etal \shortcite{mdn85}
14.4~yr earlier. This allows us to put 1-$\sigma$ upper limits
on the pulsar's proper motion of 39~mas~yr$^{-1}$ in RA and
52~mas~yr$^{-1}$ in Dec. The corresponding 1-$\sigma$ error ellipse
for the distance travelled in 1700~yr is shown in 
Figure~\ref{fig_g320_20cm}, limiting  
its birth-place to somewhere south of \rcw\ and well within the SNR.

\begin{figure*}
\begin{minipage}{160mm}
\centerline{\psfig{file=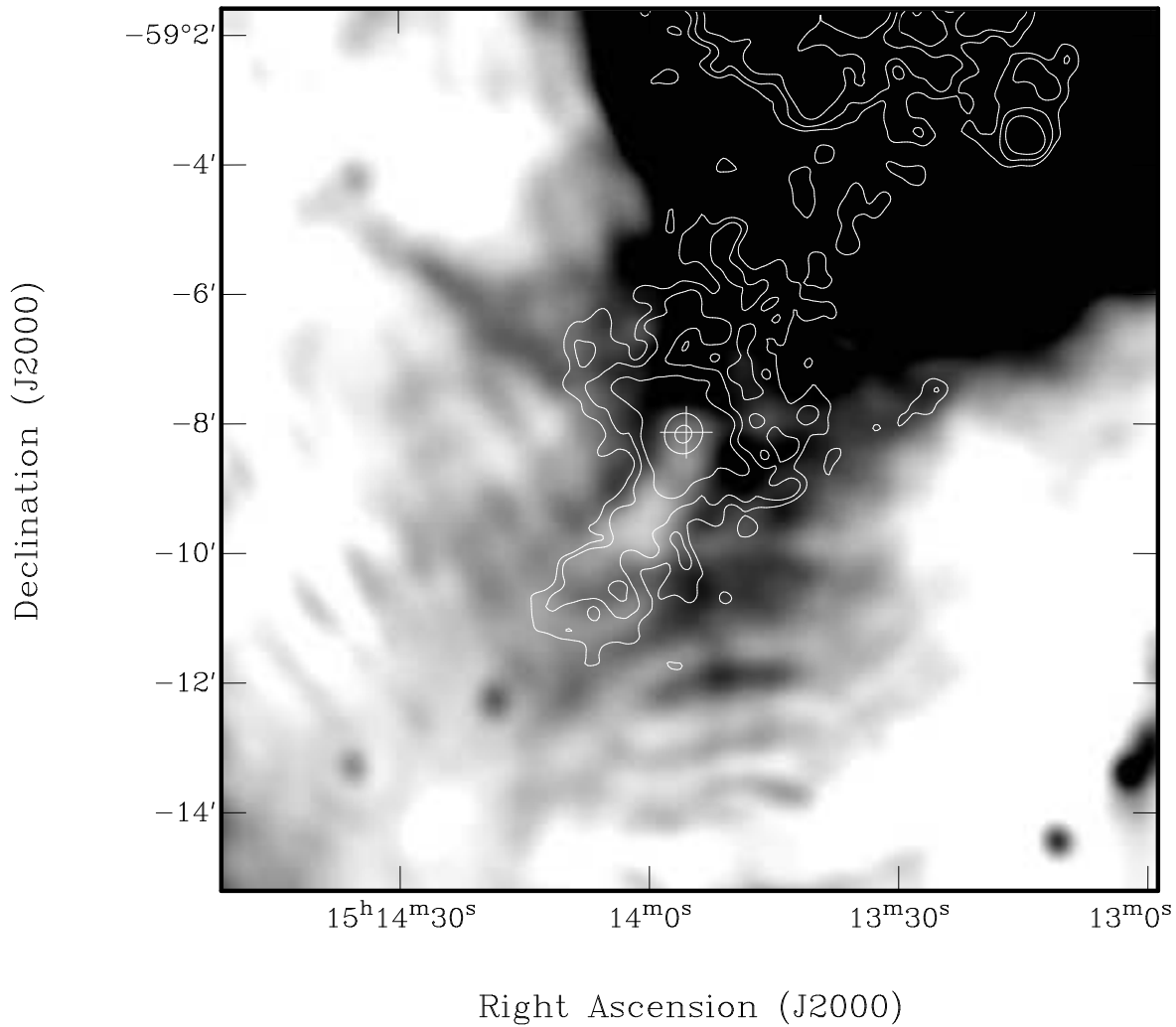,width=16cm}}
\caption{Radio/X-ray comparison of the
region near PSR~\psr.
The grey-scale corresponds to a sub-region of Figure~\ref{fig_g320_20cm},
but showing 20-cm emission in the range 1 to 10~mJy~beam$^{-1}$. 
The contours represent the {\em ROSAT}\ PSPC data of Greiveldinger \etal 
\protect\shortcite{gcm+95}, with levels (in arbitrary units) at
1.5, 2, 3, 15, 30.
Corrugations seen in the radio emission 
are low-level artifacts of the deconvolution
process.}
\label{fig_g320_pspc_2}
\end{minipage} 
\end{figure*}

An image of the region surrounding the pulsar is given in
Figure~\ref{fig_g320_pspc_2}, where the grey-scale range has been chosen to
show faint radio emission in the region. The pulsar is seen to lie within
a tongue-like plateau of emission extending 6 arcmin south-east from \rcw\
(region 6 of Figure~\ref{fig_g320_20cm}); a channel of depressed emission in
this tongue can be seen surrounding and to the south-east of the pulsar.
This tongue and the channel within it are clearly evident in MOST
observations of the same region \cite{md83,wg96}, but under conditions
of lower resolution, poorer sensitivity and reduced dynamic range. It
is also just visible in our 6-cm data, but is poorly imaged as a result
of the lack of short spacings.  X-ray emission in the region
(as imaged by {\em ROSAT}\ PSPC) is shown as contours in
Figure~\ref{fig_g320_pspc_2}.
A comparison between X-ray and radio data 
clearly shows that a narrow X-ray feature (part of the PWN powered
by PSR~\psr) runs along the channel of reduced radio emission.

\begin{table}
\caption{Observational and derived parameters for ATCA continuum data.}
\label{tab_g320_snr}
\begin{tabular}{lccc}
Wavelength (cm)  &   20   & &   6 \\ \hline
Resolution  & $24\farcs1 \times 20\farcs8$ & &
  $15\farcs0 \times 10\farcs4$ \\
Largest angular scale   & $35'$ &  & $9'$ \\
\hspace{2mm} in image \\
Measured rms noise  &  250 & (Stokes $I$) & 250  \\
\hspace{2mm} ($\mu$Jy~beam$^{-1}$)$^a$  & 75 & (Stokes $V$) & 100 \\
\\
\snr\ : \\
\hspace{2mm} Flux density (Jy)$^a$   & $28\pm1$ & (north) & $10\pm1$ \\
				 & $18\pm1$ & (south) & -- \\
				 & $46\pm2$ & (total) & -- \\
\\
PSR~\psr\ : \\
\hspace{2mm} Flux density (mJy)$^a$   & $1.3\pm0.3$ & & $<0.5$ \\
\hspace{2mm} Position (J2000)  & \multicolumn{3}{l}{RA $15^{\rm h}13^{\rm m} 
55\fs 61\pm0\fs 02$,} \\
 & \multicolumn{3}{l}{Dec $-59\degr08\arcmin08\farcs67\pm0\farcs26$} \\ 
\\
\multicolumn{2}{l}{G320.6--00.9 (Source~1 in Figure~\ref{fig_g320_20cm}) :} \\
\hspace{2mm} Flux density (Jy)$^a$   & $0.35\pm0.05$ & & $\approx0.1$ \\ 
\hspace{2mm} Position (J2000)  & \multicolumn{3}{l}{RA $15^{\rm h}15^{\rm m}
0^{\rm s}$, Dec $-58\degr46\arcmin53''$} \\ \hline
\end{tabular}

$^a$ 1 jansky (Jy)~$= 10^{-26}$~W~m$^{-2}$~Hz$^{-1}$
\end{table}

\begin{figure}
\centerline{\psfig{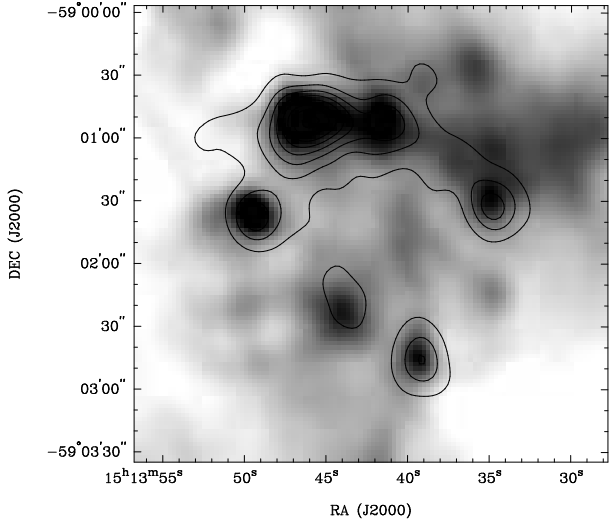}}
\caption{A high-resolution radio/X-ray comparison of the \rcw\
region. The grey-scale delineates
6-cm ATCA observations, while contours correspond to 
the {\em ROSAT} HRI data of Brazier \& Becker \protect\shortcite{bb97},
smoothed to the resolution of the 6-cm image. Contour
levels (in arbitrary units) are at levels of 5, 10, 15, $\ldots$, 40.}
\label{fig_g320_hri}
\end{figure}

In Figure~\ref{fig_g320_hri} we compare radio emission from the centre of
the \rcw\ region with the {\em ROSAT} High Resolution Imager (HRI) data
of Brazier \& Becker \protect\shortcite{bb97}.  The ring of knots seen
at the radio peak of \rcw\ shows a marked correspondence with the
X-ray emission.

To the north of the SNR
is source 1, an unresolved
core surrounded by a patchy plateau of emission of diameter $\sim$3
arcmin. This source was classified as a SNR candidate, G320.6--00.9,
by Whiteoak \& Green \shortcite{wg96}.  Its position and 20- and 6-cm
flux densities are given in Table~\ref{tab_g320_snr}.  Its flux density
at 6~cm is difficult to estimate, as the correction for the primary
beam response is large. Furthermore, 
the extended component was not imaged by the
6-cm $u-v$ coverage, and thus the corresponding flux density pertains
only to the core region.

\subsection{Polarization}
\label{sec_g320_results_pol}

\begin{figure}
\centerline{\psfig{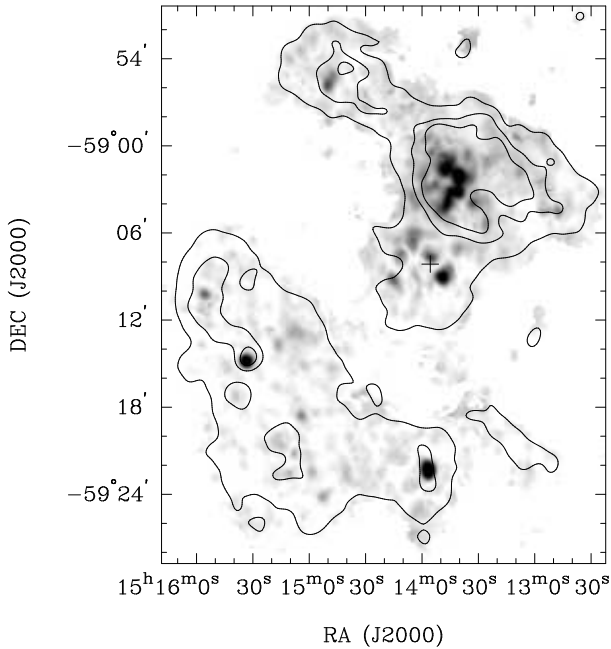}}
\caption{Polarized emission from \snr\ at 20~cm. The grey-scale runs
from 1.2 to 5.0~mJy~beam$^{-1}$. Contours represent total intensity at
20~cm smoothed to 50-arcmin resolution,
and are at levels of 20, 100 and 200~mJy~beam$^{-1}$.}
\label{fig_g320_20pol}
\end{figure}

An image of linear polarization $L$ at 20~cm is shown in
Figure~\ref{fig_g320_20pol}, showing
diffuse patchy polarization over the SNR, with
noticeable peaks in regions 4, 6, 8 and 9. Polarized intensity at
6~cm (not shown) shows similar peaks but at higher resolution. 
In Table~\ref{tab_g320_pol} we summarise the fractional polarization in
different regions, showing a general increase at 6~cm compared with 20~cm.

\begin{table}
\caption{Peak linear polarization of regions indicated
in Figure~\ref{fig_g320_20cm}.}
\label{tab_g320_pol}
\begin{tabular}{ccc}
Region &    \multicolumn{2}{c}{Fractional polarization (per cent)} \\
       &      20~cm  & 6~cm \\  \hline
1      &      $<3$  & $<3$ \\ 
2      &      17  & 40 \\
3      &       5  & 25 \\ 
4      &      10  & 18 \\
5      &      5   & 20 \\
6      &      60  & 120 \\
7      &      13  & 28 \\
8      &       7  & 30 \\
9      &      32  & --  \\ \hline
\end{tabular}
\end{table}

ATCA continuum observations involve the recording of many adjacent
frequency channels. These can be used to extract the Faraday rotation 
across the observing band, from which an accurate
rotation measure (RM) can be derived (Gaensler \etal
1998a\nocite{gmg98}).
RMs determined from
Faraday rotation across the 20-cm band are generally
low ($|RM| < 100$~rad~m$^{-2}$) except in regions 2 and 6 and on the
southern edge of region 4.  The RM of the SNR at the position of
\psr\ is 210$\pm$30~rad~m$^{-2}$, very similar to the value
of $215\pm2$~rad~m$^{-2}$ for the pulsar
(F.~Crawford 1998, private
communication). The pulsar is too weak to contaminate the
value measured for the remnant.

The fractional bandwidth at 20~cm is insufficient to accurately
determine intrinsic position angles
(see Gaensler \etal 1998a\nocite{gmg98}), while
as discussed in Section~\ref{sec_g320_reduction_cont}, 
the 6-cm $L$ image cannot be easily deconvolved.
Thus we cannot derive intrinsic position angles of linear polarization
from the SNR, and refer the reader to the lower resolution results
of Milne \etal \shortcite{mch93}. In particular, we note that
their results show that the magnetic field in the highly polarized
region 6 is oriented along the axis defined 
by the collimated X-ray feature and the associated radio channel
both seen in Figure~\ref{fig_g320_pspc_2}

The clumpy nature of polarized emission at 20~cm, along with the
significant variations in RM seen across the remnant, suggest that the low
levels of $L/I$ seen at 20~cm can be explained by beam depolarization,
in which the position angle of polarized emission varies across the
sky on scales smaller than the beam. 
As expected, the fractional polarization is
significantly higher at 6~cm, a result of both the increased resolution
and the reduced Faraday rotation at shorter wavelengths.  Since our RMs
are derived using up to 13 independent data points across the 20-cm band,
it is unlikely that there is any
ambiguity in their determination.  There is reasonable agreement between
the RMs determined here and those derived by Milne \etal
\shortcite{mch93} using only two data points, and thus one can have
confidence in the intrinsic position angles and
rotation measures which Milne \etal derive.

At 6~cm, radio emission near PSR~\psr\
(region 6) has the unphysical value of $L/I =
1.2$. This can be understood in terms of the spatial filtering applied
by an interferometer to the sky distribution of emission.  In 6-cm
total intensity, region 6 is poorly imaged by the ATCA because it is
dominated by extended structure to which the telescope is not sensitive.
But in polarization, variations in position angle (either intrinsic
to the source or resulting from differential Faraday rotation) shift
power in the polarized images into smaller-scale structure to which
the interferometer is sensitive. Polarized emission thus appears
to be of greater surface brightness than the total intensity
(see Wieringa \etal 1993\nocite{wdj+93}).  
Although the fractional polarization of region~6 is also high at 20~cm,
it does not suffer from this effect as the observations are sensitive
to all relevant spatial scales (see Section~\ref{sec_g320_results_cont}).
The high fractional polarization at 20~cm thus shows that there is minimal beam
depolarization in this region, indicating a well-ordered
magnetic field.

No linear polarization is detected from G320.6--00.9 (source~1 in
Figure~\ref{fig_g320_20cm}).
In circular polarization, the 20-cm image shows emission only from
PSR~\psr, while the 6-cm image contains no discernible sources.

\subsection{Spectral Index}
\label{sec_g320_spec_results}

Appropriate spatial filtering of the  20- and 6-cm data
results in a pair of images 
both of which contain emission on scales between 24~arcsec
(the resolution at 20~cm) and 520
arcsec (corresponding to the shortest spacings at 6~cm), and have no
structure outside this range.  
T--T plots were produced for the various regions of \snr\ marked in
Figure~\ref{fig_g320_20cm}. Several plots were produced for each region by
shifting the box in which data were considered by a few pixels each
time.  In all cases adjacent boxes produced consistent results. 

The results of the T--T fitting are shown in
Table~\ref{tab_g320_tt}.  No value of $\alpha$ could be calculated for
source 1 using this method, both because of the small number of pixels
available for the fit and because of the large uncertainties in the
primary beam correction near the edge of the field; simply
using the total flux density at 20 and 6~cm 
we estimate $\alpha^{20}_6 = -0.5\pm0.1$.  
Region 6 was only barely imaged at 6~cm because of
lack of short spacings in the array; we attempted to calculate a
spectral index for it using our 20~cm data together 
with the 36~cm image of Whiteoak \& Green \shortcite{wg96}, but its low surface
brightness and the nearness of the two wavelengths prevents us from
constraining its spectrum any more tightly than $-1 < \alpha^{36}_{20}
< 0$. Region 9 was outside the region of the 6-cm mosaic, and we
roughly estimate $\alpha^{36}_{20} = -0.5\pm0.2$.

\begin{table}
\caption{Spectral indices for regions marked in Figure~\ref{fig_g320_20cm}.}
\label{tab_g320_tt}
\begin{tabular}{cc}
Region & $\alpha^{20}_6$  \\ \hline 
1 & (see text) \\
2 & $-0.40\pm0.03$ \\ 
3 & $-0.45\pm0.04$ \\ 
4 & $-0.52\pm0.04$ \\ 
5 & $-0.42\pm0.04$ \\ 
6 & (see text) \\ 
7 & $-0.34\pm0.03$ \\ 
8 & $-0.33\pm0.03$ \\ 
9 & (see text) \\ \hline  
\end{tabular}
\end{table}

We expect the flux density scale of the ATCA images to
to be accurate to $\sim$3 per cent. While
uncertainties in absolute flux calibration will always limit the
accuracy with which $\alpha$ itself can be calculated, this effect
biases the spectral index of all regions by the same amount, and thus
does not affect our ability to comment on spatial variations in
$\alpha$.  Errors caused by applying a single primary beam correction
on images formed from a large fractional bandwidth (as is the case at
both 20 and 6~cm) can bias $\alpha$ towards the edges of the field, but
the effect is negligible ($\Delta \alpha \la 0.01$) compared to the
uncertainties quoted.

Thus results from Table~\ref{tab_g320_tt} indicate
that the spectrum of region 4 is 2--3$\sigma$ steeper than that of
regions 2 and 5, and 3--4$\sigma$ steeper than the spectra of regions 7 and ~8.

\subsection{\HI\ line}
\label{sec_g320_results_line}

One usually compares absorption spectra with emission spectra from an
adjacent patch of the sky. However, a useful emission spectrum is difficult to
extract from our ATCA data because of insufficient $u-v$ coverage on
the large spatial scales where emission dominates. We thus compare our
results to the emission spectrum in this direction obtained by Caswell
\etal \shortcite{cmr+75}. This spectrum shows significant emission at
LSR velocities of --70~\kms, --55~\kms, from --15 to +30~\kms, then at
+80~\kms.

\begin{figure*}
\begin{minipage}{160mm}
\centerline{\psfig{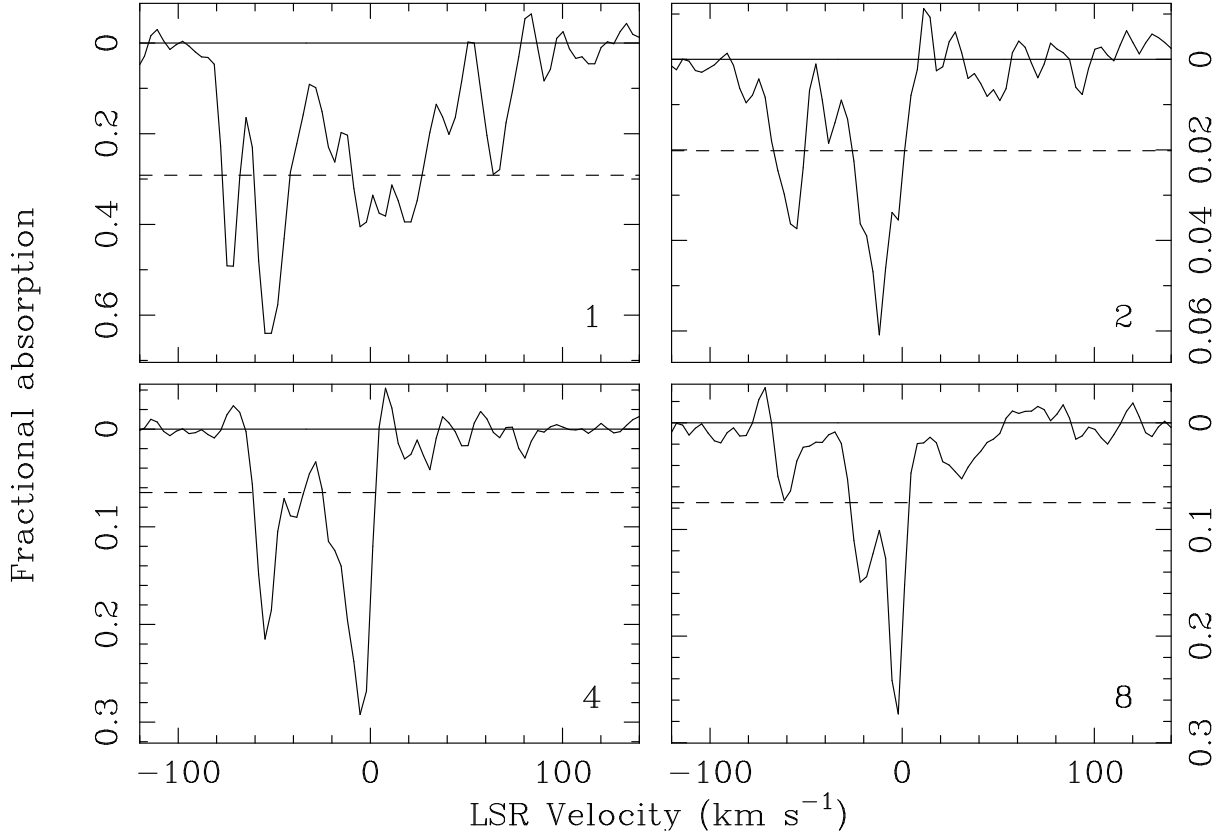}}
\caption{\HI\ absorption spectra towards four regions in the field
marked in Figure~\ref{fig_g320_20cm}. 
The dashed line marks absorption at the 6$\sigma$
level, where $\sigma$ is the noise in line-free channels.}
\label{fig_g320_hi}
\end{minipage}
\end{figure*}

Of unrelated sources near \snr, only G320.6--00.9 (source 1 in
Figure~\ref{fig_g320_20cm}) has enough signal-to-noise ratio to obtain a
useful absorption spectrum. This spectrum is shown in
Figure~\ref{fig_g320_hi}, where significant absorption is seen at --75 and
\mbox{--55~\kms}\ and between --10 and +20~\kms, with a lower level
feature at +65~\kms.  We thus put a lower limit on the systemic
velocity of G320.6--00.9 of $V_L = +20$~\kms. We use the best fitting
model for Galactic rotation of Fich, Blitz \& Stark \shortcite{fbs89},
adopt standard IAU parameters \cite{klb86} for the solar orbital
velocity ($\Theta_0 = 220$~\kms) and distance to the Galactic Centre
($R_0 = 8.5$~kpc), and assume an uncertainty of $\pm7$~\kms\ in
systemic velocities, representative of the random motion of \HI\ clouds
\cite{sra+82,bc84}. We can thus derive a lower limit on the distance to
G320.6--00.9 of $14.8\pm0.7$~kpc.

Thus G320.6--00.9 is at a large distance, is unpolarized, and has a
comparatively steep spectral index, and its core is unresolved at our
highest resolution (2~arcsec using the 6-km antenna at 6~cm). It 
therefore
seems likely that this source is not a SNR as proposed by Whiteoak \&
Green \shortcite{wg96}, but rather is a background radio galaxy. The
nature of its extended envelope is unclear.

In Figure~\ref{fig_g320_hi}
we also show \HI\ spectra towards 
three parts of \snr. Regions
2 and 4, both in the northern part of the SNR, show the same main
features in absorption: strong absorption at --55~\kms\ and then
at \mbox{--15~\kms}\ (region 2) or --5~\kms\ (region 4).
Other spectra towards the northern part of the SNR are
similar, all showing absorption out to --55~\kms. Region 8, in the
south of the SNR, shows significant absorption only at \mbox{--25~\kms}\ and
--5~\kms. A feature at --60~\kms\ is at the limit of significance.
The signal-to-noise of
absorption towards other parts of the SNR was too low to make any
useful measurements.

\subsection{OH and H140$\alpha$}
\label{sec_g320_results_oh}

Neither OH nor H140$\alpha$ emission was detected towards \snr.
The 3-$\sigma$ upper limit on the surface brightness of OH is
23~mJy~beam$^{-1}$ at 4~\kms\ resolution, 
consistent with the survey results of Green \etal
\shortcite{gfgo97} but at a slightly more stringent limit. 
Limits for H140$\alpha$ are
6~mJy~beam$^{-1}$ (15~\kms\ resolution) and 4.5~mJy~beam$^{-1}$ (40~\kms\
resolution). At the
peak of the radio emission from \rcw, the H140$\alpha$ non-detection
corresponds to an upper limit on the line-to-continuum ratio of $\sim$4~per
cent.

\section{Discussion}
\label{sec_g320_discuss}

\subsection{How many SNRs?}
\label{sec_discuss_snr_howmany}

Although there are several distinct components to the radio emission
from \snr, each part can be matched up to part of the X-ray emission in
the region.

\begin{enumerate}

\item Based on the correspondence shown in Figure~\ref{fig_g320_hri}, the
north-western radio region (which appears to be a single
object based on the similarity of \HI\ spectra
for regions 2 and 4) can be associated with
the X-ray region ``NN''.

\item The morphological correspondence seen in Figure~\ref{fig_g320_pspc_2}
between radio region 6 and the X-ray region ``SN'' argues  a connection
between them.

\item Figure~\ref{fig_g320_pspc} demonstrates a match between the ``T'' region
in X-rays and regions 7 and 8 in the radio.

\end{enumerate}

Trussoni \etal \shortcite{tmc+96} have shown that the spectra of and
absortion towards these X-ray regions are consistent with their all
being part of a single SNR.  Since each of the main radio regions
corresponds to part of this X-ray remnant, we therefore argue that {\em
the entire \snr\ region in radio and X-rays is a single SNR}.

\subsection{A distance to \snr}

Absorption spectra for regions 2 and 4 of the SNR are consistent with
that of Caswell \etal \shortcite{cmr+75}:  absorption is seen at
--55~\kms\ but not at more negative velocities. We thus adopt $V_L =
-55$~\kms, corresponding to a lower limit
on the distance of $3.8\pm0.5$~kpc. An upper limit
is the tangent point at $\sim -70$~\kms\
\cite{cmr+75,kbjk86},
which is at a kinematic distance $6.6\pm1.4$~kpc.
Region 8's spectrum
can only be considered consistent with those of regions 2 and 4 if
absorption at --60~\kms\ is genuine. However, region 8 is several arcmin
away from the other sources, and the distribution of \HI\ clouds along
the line of sight may be quite different. Indeed, values for the
hydrogen column density towards the SNR inferred from photoelectric
absorption indicate significant gradients across the SNR
\cite{tmc+96}.  Thus the \HI\ spectra can all be considered consistent 
with a distance for SNR~\snr\ of $5.2\pm1.4$~kpc.
This agrees with the
distance derived by Caswell \etal \shortcite{cmr+75}, but has more
realistic uncertainties.
In further discussion we adopt a distance to the SNR of $5d_5$~kpc. 

\subsection{Radio emission near PSR~\psr}
\label{sec_g320_discuss_snr_pwn}

Region 6 has the same RM as the pulsar (quite different from that of
\rcw), has a distinctly higher fractional polarization than \rcw, and
shows a morphological correspondence with X-ray emission associated
with the pulsar nebula.
Thus there is good evidence that region 6 is directly
associated with PSR~\psr, as originally suggested by Manchester \& Durdin
\shortcite{md83}.  Despite this claim, we would argue that this region
of radio emission is {\em not}\ a radio PWN in the traditional sense,
since the radio emission is not centred on
the pulsar, nor are the X-ray-bright regions bright in the radio. Indeed
any radio PWN is expected to be faint: comparison with the Crab Nebula
gives an expected flux density at 20~cm of $\sim$100~mJy \cite{sbd84},
while extrapolation from the X-ray spectrum of the ``SN'' region implies a
flux density $\sim$1~mJy \cite{shss84}. If we assume that this radio PWN
is of the same extent as the X-ray PWN (although in fact we expect the
radio source to be larger because of the longer synchrotron lifetimes),
we expect a 20-cm mean surface brightness at 1-arcmin resolution of
0.02--2~mJy~beam$^{-1}$, completely undetectable against the background
in this region of $\gg$20~mJy~beam$^{-1}$.

The morphology of the X-ray PWN, together with the appearance of the
surrounding SNR, have led several authors to propose that the pulsar's
wind is focused into relativistic jets or outflows, generated along an
axis aligned approximately north-west/south-east \cite{shmc83,md83,tkyb96,bb97}.
There are collimated X-ray
features along this axis on both sides of the pulsar. In the
south-east,
Figure~\ref{fig_g320_pspc_2} shows that radio emission is enhanced along
the sides of the collimated X-ray feature, but not within it.  We argue
that this anti-correspondence can be explained if the X-ray feature is
indeed a jet or outflow along the previously proposed axis; radio
emission is then produced by shocks generated in a cylindrical sheath
around the jet, as seen for SS~433 (Hjellming \& Johnston 1986,
1988\nocite{hj86,hj88}) and possibly for the X-ray jet in Vela~X
\cite{fbmo97}.  A jet interpretation is also favoured by 
the polarization properties of the sheath:
the magnetic field in the region is well-ordered
and oriented parallel to the axis of the proposed outflow
(Section~\ref{sec_g320_results_pol} above; Milne \etal 1993\nocite{mch93}),
similar to the field structure seen surrounding the jets associated
with both the Crab \cite{hss+95} and Vela \cite{mo95,mil95} pulsars.

We assume the radio sheath to be a hollow cylinder, the inner and outer
diameters of which correspond to the diameters of the X-ray jet and
radio sheath respectively. From {\em ROSAT}\ PSPC and HRI data we
estimate the inner diameter of the sheath to be 60--75~arcsec, while
from 20-cm ATCA data we determine an outer diameter of 210~arcsec.  We
assume that the jet is inclined to the line of sight at 70\degr\
\cite{bb97} and that the radio emission in the sheath is optically
thin. We then expect the radio emission along lines of sight through
the X-ray jet to be 0.7 as bright as the radio emission along the
edges of the jet, consistent with the observed value of 0.5--0.8. The
sheath is too faint for us to put any useful constraint on its spectral
index, but we expect it to have $\alpha \la -0.5$, indicative of shock
acceleration \cite{fbmo97}.

The abrupt fading of the sheath to the south-east 
is difficult to explain, but appears to
coincide with a similar termination in X-rays seen in
Figure~\ref{fig_g320_pspc_2}.
Clearly, conditions within or surrounding
the jet change in some way beyond this point.

The cross-shaped morphology of the X-ray PWN and its resemblance to the
Crab Nebula \cite{hss+95,bb97} both argue for the presence of a second
jet to the north-west of the pulsar. This jet appears to be less
collimated than the south-eastern jet and has no obvious radio
depression associated with it. While there is no clear explanation for
this difference, we note that conditions on the north and south sides
of the pulsar may be quite different (see
Section~\ref{sec_g320_discuss_finale} below), and that asymmetries are also
observed in the outflows associated with the Crab and Vela pulsars
\cite{ppp+87,mo95}.

\subsection{The knots in \rcw}
\label{sec_g320_discuss_snr_rcw}

The north-western jet which we have just discussed appears to map
onto the ring of X-ray/radio knots seen in
Figure~\ref{fig_g320_hri} \cite{bb97}. Thus rather than argue that the
entire \rcw\ region is powered by the pulsar outflow \cite{md83,tkyb96}, we
now consider the alternative that the interaction region is confined
just to the ring of knots.

The radio knots within \rcw\ are significantly linearly
polarized, show no recombination lines and have the
steepest spectral index of any emission from the SNR. Thus there seems
little doubt that the emission mechanism associated with the radio
features is
synchrotron radiation. Taking the brightest knot 
to be a sphere of radius 15~arcsec and flux
density 0.2~Jy at 20~cm, we can use standard minimum energy arguments
\cite{pac70} to infer a magnetic field for it of $1.5(1+k)^{2/7}
\phi^{-2/7} d_5^{-2/7}$~$\mu$G, where $\phi$ is the filling factor of
emitting particles and fields and $k$ is the ratio of energy in heavy
particles to that in electrons. This is not significantly different
from the field strengths of 7~$\mu$G and 2.5~$\mu$G inferred for the
overall X-ray PWN \cite{shss84} and for the X-ray jet \cite{tkyb96}
respectively. The corresponding total energy of this knot in magnetic fields
and relativistic particles is $1.3 (1+k)^{4/7}\phi^{3/7}d_5^{17/7} 
\times10^{42}$~erg, a tiny fraction of the total spin-down energy
released by the pulsar in its life-time ($>10^{48}$~erg).
The steep spectrum of the knots is the opposite to what we might expect
if injection of particles from the pulsar was directly contributing to
their emission (cf. \ Frail \etal 1994b\nocite{fkw94}). Rather, it
seems that the collision between the jet and pre-existing material
drives shocks into the latter at the working surface, producing
clumps of enhanced synchrotron emission.

X-ray observations of \rcw\ as a whole show it to be thermal
\cite{shss84,tkyb96}, but the X-ray/radio correspondence seen in 
Figure~\ref{fig_g320_hri} suggests that X-ray emission from the knots
might be synchrotron emission as in the radio. A single power law of
spectral index $-0.6 \la \alpha \la -0.5$ (from
Table~\ref{tab_g320_tt}) can indeed be extrapolated between the radio
flux of the knots as observed here and the X-ray count rates quoted by
Brazier \& Becker \shortcite{bb97}. No spectral break as might be
caused by synchtrotron losses is required, which implies a magnetic
field strength $<10$~$\mu$G consistent with the determination from
minimum energy above.  While \rcw\ is predominantly thermal in X-rays,
the available spectral data lack the spatial resolution to determine
whether the thermal spectrum of the region corresponds to the knots as
well as the more extended component in which they are embedded.
Comparison of the count-rate for the entire \rcw\ region \cite{tmc+96}
with those of the knots alone \cite{bb97} shows that soft X-ray
emission is dominated by the extended component, and thus at low
energies a power-law spectrum associated with the knots might be
hidden.  Indeed observations in hard X-rays \cite{tkyb96} show
non-thermal emission at their position. However, the spatial resolution
of these data is not sufficient to separate the knots from more
diffuse emission in the area; high-resolution observations with {\em
AXAF}\ will be required to clarify the issue. Whether thermal or
non-thermal, the energy in the knots is far smaller than that in the
whole \rcw\ region, and so can be
comfortably accounted for by the pulsar spin-down.

The horseshoe-shaped plateau in \rcw\ appears to have a similar shape
to the ring it envelopes, and may represent material diffusing away
from the point(s) of impact. However, the size of this region is then
somewhat large:  from the ring to the edge of this plateau region is
2.5~arcmin, which requires emission to have travelled outwards from the
ring at a projected mean velocity of $2100d_5$~\kms. One possibility
suggested by Tamura \etal \shortcite{tkyb96} is that the pulsar jet has
precessed or narrowed in opening angle.

To summarise, the problems raised at the end of
Section~\ref{sec_g320_intro_questions_assoc}
can be resolved if just the
X-ray/radio knots, rather than the whole of \rcw, are interpreted as
the point of interaction between the pulsar outflow and the SNR.  The
energy in these knots can be easily accomodated by the pulsar
spin-down, and it is easier to understand how these features,
rather than the whole extended nebula, might be
the termination of an energetic outflow.
A more likely possibility to account for the surrounding
\rcw\ region is that it originates from the SNR blast wave, and is not
related to any outflow from the pulsar.

\subsection{The south-eastern component of the SNR}
\label{sec_g320_discuss_snr_southeast}

\begin{figure}
\centerline{\psfig{file=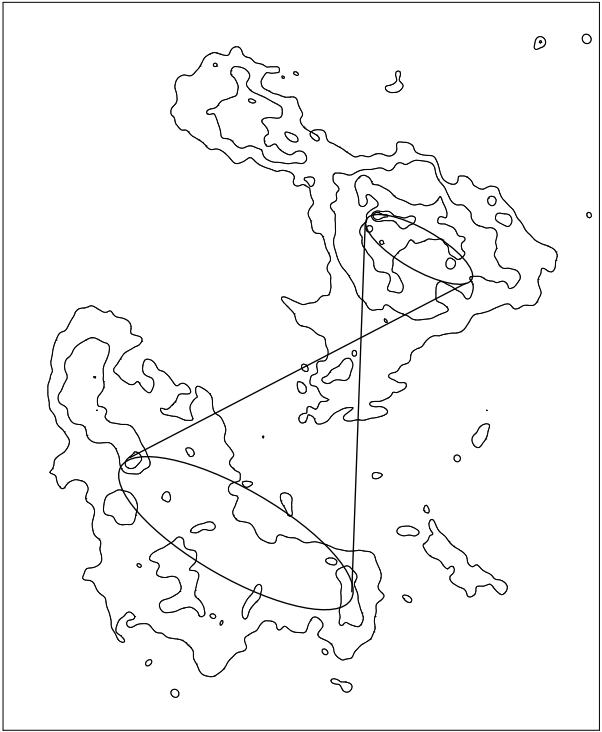,width=8cm,angle=270}}
\caption{Speculative geometry for outflow in \snr. Contours
represent 20-cm continuumm, and are at 5, 20, 60, 120 and
150~mJy~beam$^{-1}$.}
\label{fig_g320_20con}
\end{figure}

It can be seen in Figure~\ref{fig_g320_20pol}
that regions 8 and 9 of the SNR contain compact polarized clumps, resembling
the knots seen in \rcw.  These features may be evidence for an
interaction between the south-eastern jet and the south-eastern half of
the SNR, although, 
compared to that with the \rcw\ region, the impact is 
greatly reduced.
In Figure~\ref{fig_g320_20con} we show an
extension of the model proposed by Brazier \& Becker \shortcite{bb97},
where we speculatively project an additional cone of emission onto the
south-eastern part of the SNR.\footnote{This projection is qualitatively
similar to that of Manchester \& Durdin \shortcite{md83} and Manchester
\shortcite{man87}, but maps onto a smaller region of each half of the
SNR.}  It is interesting to note that this second cone happens to
intersect the SNR at the location of these two highly polarized
clumps.  It can be seen in Figure~\ref{fig_g320_pspc} that while the X-ray
emission joining the ``SN'' and ``T'' regions might make up one
edge of this
cone, the other edge is missing. However, the latter region
lies right along a rib of the PSPC and so may have had minimal
exposure.

\subsection{Spectral index variations}

Results presented in Section~\ref{sec_g320_spec_results} indicate that the
compact knots at the peak of \rcw\ have a steeper radio spectrum than the
diffuse emission surrounding them, which are again steeper than the
south-eastern part of the SNR.  While the phenomenon of spatial spectral
index variations in SNRs is not well understood, there is the suggestion that
young remnants generally show a steeper spectrum in their brightest, most
compact components \cite{ar93}, as observed here.  For example, in
G260.4--03.4 (Puppis~A) the steepest spectrum part of the shell is where
it has been apparently decelerated by a collision with a dense cloud
\cite{dbwg91}, while in G111.7--02.1 (Cassiopeia~A), the steep spectrum
regions tend to be bow-shocks driven by clumps of fast-moving material
\cite{arl+91}.  We note that both these effects may be occurring
in \snr: significant deceleration of the \rcw\ region is argued for in
Section~\ref{sec_g320_discuss_finale} below, while the pulsar outflow may
accelerate small clumps of material to high velocity, which then drive
bow-shocks as in Cas~A.

\subsection{Finale: what is SNR~\snr?}
\label{sec_g320_discuss_finale}

The distance to the SNR of 4.2~kpc as derived by Caswell \etal\
\shortcite{cmr+75} was used in the calibration of the electron
distribution model of Taylor \& Cordes \shortcite{tc93}. Thus
comparisons between the pulsar's distance as derived from this
model and any distance estimate to the SNR are not
entirely meaningful.  Rather, we estimate the pulsar's distance 
as follows.

While PSR~\psr\ itself is too weak to show \HI\ absorption,
PSRs B1240--64, B1323--58, B1323--62, B1356--60 and B1557--50 
are all within 20\degr\ of PSR~\psr, have comparable dispersion
measures and have independent distance estimates from
\HI\ absorption \cite{fw90,sdw+96}. Distances and dispersion measures
for these pulsars suggest a mean electron density along
this line of sight of between 0.03 and 0.06~cm$^{-3}$, corresponding
to a distance for PSR~\psr\ in the range 4.2 to 8.4~kpc.

This distance for the pulsar is consistent
with the distance we have determined for SNR~\snr\ through \HI\
absorption, resolving any previous discrepancy.
Furthermore, despite the large
variation in RM across the SNR (--200 to +400~rad~m$^{-2}$), its RM at
the position of the pulsar agrees with that of the pulsar
itself. Finally, we discussed in Section~\ref{sec_g320_discuss_snr_rcw}
that the pulsar appears to be interacting with the
surrounding SNR. Thus we argue strongly that {\em
PSR~\psr\ and SNR~\snr\ are associated}. 
The age discrepancy outlined
in Section~\ref{sec_g320_intro_questions_assoc} must then be resolved.

A first possibility to account for this apparent age difference is that
\psr\ has had an unusual spin-down history, and is much older
than it seems \cite{br88}. However, all observations
suggest that \psr\ is a typical young pulsar, but that \snr\ is an
anything-but-typical remnant. Although the possibility that \psr\ is
old cannot be ruled out, disagreement between a relatively well-determined
pulsar age and a crudely estimated SNR age gives little cause to look
to the pulsar as the culprit. Thus in future discussion
we assume that the
age of the system is $\la$1700~yr. We now attempt to account for this
age in terms of the properties of the supernova and its environment.

We take the south-eastern component of the SNR (incorporating regions
7, 8 and 9) as that part most representative of a ``normal'' shell
SNR.  Assuming the site of the supernova explosion to be within the
1-$\sigma$ error-box derived from the pulsar's proper motion, we
estimate the radius of the south-eastern component 
of the SNR to be $17\pm2$~arcmin.  For an
SNR age of 1700~yr, this implies a mean expansion velocity of
$(14\pm2)d_5 \times 10^3$~\kms. If this component represents freely
expanding material, we find a kinetic energy for the explosion of
$E_{51} = (2.0\pm0.3)M_{\rm ej}d_5^2$, where $M_{\rm ej}$ is the
ejected mass in units of M$_{\sun}$ and
$E_{51}$ is in units of $10^{51}$~erg.  Alternatively,
if we assume expansion in the adiabatic (Sedov-Taylor) phase then the size
of the SNR implies $(E_{51}/n_1) \sim (1100\pm600)d_5^5$, where
$n_1$ is the ambient density into which this component of the
SNR is expanding.  To fully
enter this phase the SNR must sweep up 20 times its own mass
\cite{fbs83,dj96}, implying a density
$n_1 > (0.012\pm0.003)M_{\rm ej}d_5^{-3}$~cm$^{-3}$ to give $E_{51}
\ga (3.3\pm2.9)M_{\rm ej}d_5^2$.  

Thus regardless of the evolutionary state of the SNR, an age of 1700~yr
implies a value of $E_{51}/M_{\rm ej}$ which is 10--100 times higher
than typical values \cite{ber88,smi88b}. This problem has been overcome
in the past by invoking a large ($E_{51} \sim 10-100$) kinetic energy
for the supernova \cite{shmc83,kat83}. While such a value for $E_{51}$
seems uncomfortably high, it is comparable with (model-dependent) estimates
for the energy released in
the recent Type~Ic supernova SN~1998bw \cite{imn+98,wes98}. However, we
propose the more likely explanation that the progenitor of
PSR~\psr\ was a helium star. Such a star explodes in a Type~Ib
supernova, ejecting a low amount of mass ($M_{\rm ej} \sim 1.5$) but
with a typical supernova energy ($E_{51} \sim 1-2$) \cite{wlw95}, as
required here.  While the optical filaments in RCW~89 show no evidence
of helium rich ejecta \cite{dmf77,shmc83}, we show below that this
component of the SNR is considerably evolved and is now dominated by
swept-up material.  Spectroscopy of any optical emission which can be
associated with the south-eastern component of \snr\ would be of great
interest.

Helium stars start off as extremely massive stars, but lose their
hydrogen envelopes either through a strong stellar wind or through
transer of mass to a binary companion (Woosley, Langer \& Weaver 1993,
1995\nocite{wlw93,wlw95}). The latter possibility is interesting,
particularly as the O~star Muzzio~10 \cite{muz79,shmc83} is just
18~arcsec north of the pulsar, and is at a comparable distance
($\sim$5~kpc; Arendt~1991\nocite{are91}).  We thus speculate that Muzzio~10
is the former companion of PSR~\psr. This possibility is easily
tested:  assuming a pulsar age of 1700~yr, the association predicts a
proper motion for the pulsar of 11~mas~yr$^{-1}$ along a position angle
168\degr\ (N through E), corresponding to a transverse velocity
260$d_5$~\kms.

It is clear from the radio data that there is no
radio emission which can trace an entire shell and thus join
the two components of the SNR.  Gaensler \shortcite{gae98} has argued
that the ``bilateral'' or ``barrel'' morphology which results (and the
observed alignment of the bilateral axis with the Galactic Plane) is
produced when a SNR expands into an elongated low-density cavity,
radio emission being produced where the shock has recently
encountered the walls. Thus a bilateral appearance can be considered
evidence that $n_1$ is low ($\sim0.01$~cm$^{-3}$); such a value for
$n_1$ can also account for the faintness of the radio PWN \cite{bha90},
and is consistent with the cavities expected due to 
the presence of other massive stars
in the region (Lortet et al. 1987\nocite{lgg87}). For $M_{\rm ej} = 1.5$,
this ambient density corresponds to the south-eastern component
beginning to enter the adiabatic phase. 

Limits on the pulsar proper motion as indicated in
Figure~\ref{fig_g320_20cm} demonstrate that the SNR is distributed highly
asymmetrically about the site of the explosion. Intuitively, this
suggests that the supernova explosion was near the edge
of the proposed cavity, and that, while the south-eastern half of the SNR
has expanded relatively unimpeded, the north-western half has
decelerated significantly.  A high-density environment for this
component is also argued for by its high X-ray and radio brightness and
by the small proper motion of associated optical filaments
\cite{shmc83,man92b}.

Gull \shortcite{gul73} has modelled the
deceleration of a young SNR, and provides means of scaling his solution
to arbitrary values of density and $E_{51}/M_{\rm ej}$. We
take the radius of the north-western component to be $7\pm2$~arcmin,
and adopt $E_{51}/M_{\rm ej} \sim 2$ as argued for above.
Scaling to Figure~3 of Gull
\shortcite{gul73}, we find that the north-western component of the SNR
is well into the adiabatic phase, and that the ambient density
in this region is in the range $n_2 = 1 - 5$~cm$^{-3}$. This is
consistent with the non-detection of OH, as the 1720 MHz maser line
will be shock-excited into emission only at much higher densities
\cite{eli76}.

To summarise, we find that a number of observed properties of the SNR,
particularly its bilateral appearance, faint radio PWN, asymmetric
distribution about the pulsar and large size (and possibly also the
slow optical expansion of filaments in RCW~89) can all be explained in
a model in which an explosion of high energy or low mass occurred near the
edge of a cavity elongated parallel to the Galactic Plane. The density
of the cavity is $\sim0.01$~cm$^{-3}$: the south-eastern component of
the SNR has expanded rapidly across the cavity and has recently
collided with the other side, while to the north-west, the shock has
encountered denser (1--5~cm$^{-3}$) material, causing the SNR to
decelerate significantly.  To the north-east and south-west the shock
is still propagating through the cavity and produces no observable
emission.

While at low resolution G320.6--01.6 appears to be an extension of
\snr\ (Milne et al. 1993\nocite{mch93}), 
it is impossibly large to have expanded to its
current size in 1700~yr. Its morphology and surface brightness are
in complete contrast to those of \snr, and we conclude that it is
an unrelated older SNR along the same line of sight.

\section{Conclusion}
\label{sec_g320_conclusion}

We have reported on an extensive set of 
ATCA observations of \snr\ and PSR~\psr. The main results
of this study are as follows.

\begin{enumerate}

\item The disparate radio components of \snr\ are all part of a single SNR
at a distance of $5.2\pm1.4$~kpc and with an age of $\sim$1700~yr.

\item PSR~\psr\ is physically associated with \snr. The pulsar emits
twin jets or collimated outflows of relativistic particles, one of
which interacts with the SNR in the form of radio/X-ray knots within
\rcw.

\item SNR~\snr\ was formed in a supernova of high
kinetic energy or low ejected mass ($E_{51}/M_{\rm ej} \sim 2$)
which occurred near the edge of a low-density cavity.
\end{enumerate}

A variety of further observations can support or refute these conclusions.
For example, the cavity we propose may be visible in \HI\ emission, 
while forthcoming {\em AXAF}\ observations 
will allow a detailed study of the physical
conditions within the knots.

While we have offered answers to the questions outlined in
Section~\ref{sec_g320_intro}, the picture is far from complete.  First, we
lack an understanding of the physical details of how the pulsar outflow
is generating the peculiar knots in \rcw, and of how the pulsar is
interacting with the south-eastern half of the SNR.  Similar
interactions through outflows or jets have been proposed for a variety
of other SNRs \cite{rmk+85,man87,ggm98}, although an associated pulsar
is yet to be detected in most cases.  We have also sidestepped the
whole issue of why PSR~\psr\ generates such large-scale jets and how
general this situation might be. Certainly the theoretical expectation
has been for some time that pulsars should generate collimated outflows
\cite{ben84,mic85,sl90}. The Crab, Vela, PSR~\psr\ and possibly
PSR~B1951+32 (Hester 1998\nocite{hes98a}; J. J. Hester et al. in preparation)
all show evidence that
this is indeed the case, 
while various jet-like features have been proposed around
other pulsars \cite{bel97}. 
We must now therefore accept that spherically
symmetric pulsar winds are a gross over-simplification.

\section*{Acknowledgments}

We thank
Douglas Bock for carrying out the OH observations, Roy Duncan
for obtaining Parkes data on this source and Warwick Wilson for
assistance with pulsar-gating. We also appreciate useful discussions
with Douglas Bock, John Dickey, Richard Dodson, Neil Killeen,
Sne\v{z}ana Stanimirovi\'{c} and Ben Stappers.  Chris Greiveldinger and
Silvano Massaglia provided X-ray images of the region, while Froney
Crawford communicated the rotation measure of PSR~\psr\ prior to
publication.  BMG acknowledges the support of an Australian
Postgraduate Award.  The Australia Telescope is funded by the
Commonwealth of Australia for operation as a National Facility managed
by CSIRO.  This research has made use of NASA's Astrophysics Data
System Abstract Service and of the SIMBAD database, operated at CDS,
Strasbourg, France.

\bibliographystyle{mn}
\bibliography{modrefs,psrrefs}

\begin{thebibliography}{{Srinivasan, Bhattacharya \& Dwarakanath }{1984}}

\bibitem[\protect\citename{Anderson \& Rudnick }{1993}]{ar93}
Anderson~M.~C., Rudnick~L., 1993, ApJ, 408, 514

\bibitem[\protect\citename{Anderson {\rm et~al. }}{1991}]{arl+91}
Anderson~M., Rudnick~L., Leppik~P., Perley~R., Braun~R., 1991, ApJ, 373, 146

\bibitem[\protect\citename{Arendt }{1991}]{are91}
Arendt~R.~G., 1991, AJ, 101, 2160

\bibitem[\protect\citename{Belfort \& Crovisier }{1984}]{bc84}
Belfort~P., Crovisier~J., 1984, AA, 136, 368

\bibitem[\protect\citename{Bell }{1997}]{bel97}
Bell~J.~F., 1997, Vistas Astron., 41, 87

\bibitem[\protect\citename{Benford }{1984}]{ben84}
Benford~G., 1984, ApJ, 282, 154

\bibitem[\protect\citename{Berkhuijsen }{1988}]{ber88}
Berkhuijsen~E.~M., 1988, AA, 192, 299

\bibitem[\protect\citename{Bhattacharya }{1990}]{bha90}
Bhattacharya~D., 1990, JA\&A, 11, 125

\bibitem[\protect\citename{Blandford \& Romani }{1988}]{br88}
Blandford~R.~D., Romani~R.~W., 1988, MNRAS, 234, 57P

\bibitem[\protect\citename{Braun, Goss \& Lyne }{1989}]{bgl89}
Braun~R., Goss~W.~M., Lyne~A.~G., 1989, ApJ, 340, 355

\bibitem[\protect\citename{Brazier \& Becker }{1997}]{bb97}
Brazier~K. T.~S., Becker~W., 1997, MNRAS, 284, 335

\bibitem[\protect\citename{Caswell, Milne \& Wellington }{1981}]{cmw81}
Caswell~J.~L., Milne~D.~K., Wellington~K.~J., 1981, MNRAS, 195, 89

\bibitem[\protect\citename{Caswell {\rm et~al. }}{1975}]{cmr+75}
Caswell~J.~L., Murray~J.~D., Roger~R.~S., Cole~D.~J., Cooke~D.~J., 1975, AA,
  45, 239

\bibitem[\protect\citename{Clark }{1980}]{cla80}
Clark~B.~G., 1980, AA, 89, 377

\bibitem[\protect\citename{Cornwell }{1988}]{cor88}
Cornwell~T.~J., 1988, AA, 202, 316

\bibitem[\protect\citename{Day, Thomas \& Goss }{1969}]{dtg69}
Day~G.~A., Thomas~B.~M., Goss~W.~M., 1969, Aust.\,J.\,Phys.\,Astr.\,Supp., 11,
  11

\bibitem[\protect\citename{Dickey }{1997}]{dic97}
Dickey~J.~M., 1997, ApJ, 488, 258

\bibitem[\protect\citename{Dohm-Palmer \& Jones }{1996}]{dj96}
Dohm-Palmer~R.~C., Jones~T.~W., 1996, ApJ, 471, 279

\bibitem[\protect\citename{Dopita, Mathewson \& Ford }{1977}]{dmf77}
Dopita~M.~A., Mathewson~D.~S., Ford~V.~L., 1977, ApJ, 214, 179

\bibitem[\protect\citename{du~Plessis {\rm et~al. }}{1995}]{ddb+95}
du~Plessis~I., de~Jager~O.~C., Buchner~S., Nel~H.~I., North~A.~R.,
  Raubenheimer~B.~C., van~der Walt~D.~J., 1995, ApJ, 453, 746

\bibitem[\protect\citename{Dubner {\rm et~al. }}{1991}]{dbwg91}
Dubner~G.~M., Braun~R., Winkler~P.~F., Goss~W.~M., 1991, AJ, 101, 1466

\bibitem[\protect\citename{Elitzur }{1976}]{eli76}
Elitzur~M., 1976, ApJ, 203, 124

\bibitem[\protect\citename{Fabian, Brinkmann \& Stewart }{1983}]{fbs83}
Fabian~A.~C., Brinkmann~W., Stewart~G.~C., 1983, in Danziger~J., Gorenstein~P.,
  eds, Supernova Remnants and Their X-Ray Emission ({IAU} {S}ymposium 101).
\newblock Reidel, Dordrecht, p.~119

\bibitem[\protect\citename{Fich, Blitz \& Stark }{1989}]{fbs89}
Fich~M., Blitz~L., Stark~A.~A., 1989, ApJ, 342, 272

\bibitem[\protect\citename{Frail \& Weisberg }{1990}]{fw90}
Frail~D.~A., Weisberg~J.~M., 1990, AJ, 100, 743

\bibitem[\protect\citename{Frail, Kassim \& Weiler }{1994}]{fkw94}
Frail~D.~A., Kassim~N.~E., Weiler~K.~W., 1994, AJ, 107, 1120

\bibitem[\protect\citename{Frail {\rm et~al. }}{1997}]{fbmo97}
Frail~D.~A., Bietenholz~M.~F., Markwardt~C.~B., \"{O}gelman~H., 1997, ApJ, 475,
  224

\bibitem[\protect\citename{Frater, Brooks \& Whiteoak }{1992}]{fbw92}
Frater~R.~H., Brooks~J.~W., Whiteoak~J.~B., 1992,
  J.\,Electr.\,Electron.\,Eng.\,Aust., 12, 103

\bibitem[\protect\citename{Gaensler }{1998}]{gae98}
Gaensler~B.~M., 1998, ApJ, 493, 781

\bibitem[\protect\citename{Gaensler \& Johnston }{1995a}]{gj95b}
Gaensler~B.~M., Johnston~S., 1995a, Publ.\,Astron.\,Soc.\,Aust., 12, 76

\bibitem[\protect\citename{Gaensler \& Johnston }{1995b}]{gj95c}
Gaensler~B.~M., Johnston~S., 1995b, MNRAS, 277, 1243

\bibitem[\protect\citename{Gaensler, Manchester \& Green }{1998a}]{gmg98}
Gaensler~B.~M., Manchester~R.~N., Green~A.~J., 1998a, MNRAS, 296, 813

\bibitem[\protect\citename{Gaensler, Green \& Manchester }{1998b}]{ggm98}
Gaensler~B.~M., Green~A.~J., Manchester~R.~N., 1998b, MNRAS, 299, 812

\bibitem[\protect\citename{Green }{1990}]{gre90}
Green~D.~A., 1990, AJ, 100, 1927

\bibitem[\protect\citename{Green {\rm et~al. }}{1997}]{gfgo97}
Green~A.~J., Frail~D.~A., Goss~W.~M., Otrupcek~R., 1997, AJ, 114, 2058

\bibitem[\protect\citename{Greiveldinger {\rm et~al. }}{1995}]{gcm+95}
Greiveldinger~C., Caucino~S., Massaglia~S., \"{O}gelman~H., Trussoni~E., 1995,
  ApJ, 454, 855

\bibitem[\protect\citename{Gull \& Daniell }{1978}]{gd78}
Gull~S.~F., Daniell~G.~J., 1978, Nat, 272, 686

\bibitem[\protect\citename{Gull }{1973}]{gul73}
Gull~S.~F., 1973, MNRAS, 161, 47

\bibitem[\protect\citename{Hester }{1998}]{hes98a}
Hester~J.~J., 1998, in Shibazaki~N., Kawai~N., Shibata~S., Kifune~T., eds,
  Neutron Stars and Pulsars: Thirty Years after the Discovery.
\newblock Universal Academy Press, Tokyo, p.~431

\bibitem[\protect\citename{Hester {\rm et~al. }}{1995}]{hss+95}
Hester~J.~J. {\rm et~al.}, 1995, ApJ, 448, 240

\bibitem[\protect\citename{Hill }{1968}]{hil68}
Hill~E.~R., 1968, Aust.\,J.\,Phys., 21, 735

\bibitem[\protect\citename{Hjellming \& Johnston }{1986}]{hj86}
Hjellming~R.~M., Johnston~K.~J., 1986, in Mason~K.~O., Watson~M.~G.,
  White~N.~E., eds, {The Physics of Accretion onto Compact Objects}.
\newblock Springer-Verlag, Berlin, p.~287

\bibitem[\protect\citename{Hjellming \& Johnston }{1988}]{hj88}
Hjellming~R.~M., Johnston~K.~J., 1988, ApJ, 328, 600

\bibitem[\protect\citename{Iwamoto {\rm et~al. }}{1998}]{imn+98}
Iwamoto~K. {\rm et~al.}, 1998, Nat, in press (astro-ph/9806382)

\bibitem[\protect\citename{Kafatos {\rm et~al. }}{1980}]{ksbg80}
Kafatos~M., Sofia~S., Bruhweiler~F., Gull~S., 1980, ApJ, 242, 294

\bibitem[\protect\citename{Kaspi }{1996}]{kas96}
Kaspi~V.~M., 1996, in Johnston~S., Walker~M.~A., Bailes~M., eds, Pulsars:
  Problems and Progress ({IAU} Colloquium 160).
\newblock Astronomical Society of the Pacific, San Francisco, p.~375

\bibitem[\protect\citename{Kaspi {\rm et~al. }}{1994}]{kms+94}
Kaspi~V.~M., Manchester~R.~N., Siegman~B., Johnston~S., Lyne~A.~G., 1994, ApJ,
  422, L83

\bibitem[\protect\citename{Katz }{1983}]{kat83}
Katz~J.~I., 1983, AA, 128, L1

\bibitem[\protect\citename{Kerr \& Lynden-Bell }{1986}]{klb86}
Kerr~F.~J., Lynden-Bell~D., 1986, MNRAS, 221, 1023

\bibitem[\protect\citename{Kerr {\rm et~al. }}{1986}]{kbjk86}
Kerr~F.~J., Bowers~P.~F., Jackson~P.~D., Kerr~M., 1986, A\&AS, 66, 373

\bibitem[\protect\citename{Kesteven }{1968}]{kes68}
Kesteven~M. J.~L., 1968, Aust.\,J.\,Phys., 21, 369

\bibitem[\protect\citename{Komesaroff }{1966}]{kom66}
Komesaroff~M.~M., 1966, Aust.\,J.\,Phys., 19, 75

\bibitem[\protect\citename{Lortet, Georgelin \& Georgelin }{1987}]{lgg87}
Lortet~M.~C., Georgelin~Y.~P., Georgelin~Y.~M., 1987, AA, 180, 65

\bibitem[\protect\citename{Manchester \& Durdin }{1983}]{md83}
Manchester~R.~N., Durdin~J.~M., 1983, in Danziger~J., Gorenstein~P., eds,
  Supernova Remnants and Their {X}-Ray Emission ({IAU} {S}ymposium 101).
\newblock Reidel, Dordrecht, p.~421

\bibitem[\protect\citename{Manchester }{1987}]{man87}
Manchester~R.~N., 1987, AA, 171, 205

\bibitem[\protect\citename{Manchester }{1992}]{man92b}
Manchester~R.~N., 1992, Nat, 356, 660

\bibitem[\protect\citename{Manchester, Durdin \& Newton }{1985}]{mdn85}
Manchester~R.~N., Durdin~J.~M., Newton~L.~M., 1985, Nat, 313, 374

\bibitem[\protect\citename{Manchester, Tuohy \& D'Amico }{1982}]{mtd82}
Manchester~R.~N., Tuohy~I.~R., D'Amico~N., 1982, ApJ, 262, L31

\bibitem[\protect\citename{Markwardt \& \"{O}gelman }{1995}]{mo95}
Markwardt~C.~B., \"{O}gelman~H., 1995, Nat, 375, 40

\bibitem[\protect\citename{Marsden {\rm et~al. }}{1997}]{mbg+97}
Marsden~D. {\rm et~al.}, 1997, ApJ, 491, L39

\bibitem[\protect\citename{Michel }{1985}]{mic85}
Michel~F.~C., 1985, ApJ, 288, 138

\bibitem[\protect\citename{Mills, Slee \& Hill }{1960}]{msh60}
Mills~B.~Y., Slee~O.~B., Hill~E.~R., 1960, Aust.\,J.\,Phys., 13, 676

\bibitem[\protect\citename{Milne \& Dickel }{1975}]{md75}
Milne~D.~K., Dickel~J.~R., 1975, Aust.\,J.\,Phys., 28, 209

\bibitem[\protect\citename{Milne }{1970}]{mil70}
Milne~D.~K., 1970, Aust.\,J.\,Phys., 23, 425

\bibitem[\protect\citename{Milne }{1972}]{mil72b}
Milne~D.~K., 1972, Aust.\,J.\,Phys., 25, 307

\bibitem[\protect\citename{Milne }{1995}]{mil95}
Milne~D.~K., 1995, MNRAS, 277, 1435

\bibitem[\protect\citename{Milne, Caswell \& Haynes }{1993}]{mch93}
Milne~D.~K., Caswell~J.~L., Haynes~R.~F., 1993, MNRAS, 264, 853

\bibitem[\protect\citename{Milne {\rm et~al. }}{1969}]{mwgm69}
Milne~D.~K., Wilson~T.~L., Gardner~F.~F., Mezger~P.~G., 1969,
  Astrophys.\,Lett., 4, 121

\bibitem[\protect\citename{Muzzio }{1979}]{muz79}
Muzzio~J.~C., 1979, AJ, 84, 639

\bibitem[\protect\citename{Pacholczyk }{1970}]{pac70}
Pacholczyk~A.~G., 1970, Radio Astrophysics.
\newblock Freeman, San Francisco

\bibitem[\protect\citename{Pelling {\rm et~al. }}{1987}]{ppp+87}
Pelling~R.~M., Paciesas~W.~S., Peterson~L.~E., Makishima~K., Oda~M.,
  Ogawara~Y., Miyamoto~S., 1987, ApJ, 319, 416

\bibitem[\protect\citename{Reynolds }{1994}]{rey94}
Reynolds~J.~E., 1994, ATNF Technical Document Series, 39.3040

\bibitem[\protect\citename{Rodgers, Campbell \& Whiteoak }{1960}]{rcw60}
Rodgers~A.~W., Campbell~C.~T., Whiteoak~J.~B., 1960, MNRAS, 121, 103

\bibitem[\protect\citename{Roger {\rm et~al. }}{1985}]{rmk+85}
Roger~R.~S., Milne~D.~K., Kesteven~M.~J., Haynes~R.~F., Wellington~K.~J., 1985,
  Nat, 316, 44

\bibitem[\protect\citename{Saravanan {\rm et~al. }}{1996}]{sdw+96}
Saravanan~T.~P., Deshpande~A.~A., Wilson~W., Davies~E., McCulloch~P.~M.,
  McConnell~D., 1996, MNRAS, 280, 1027

\bibitem[\protect\citename{Sault \& Killeen }{1998}]{sk97}
Sault~R.~J., Killeen~N. E.~B., 1998, The {\tt MIRIAD} User's Guide.
\newblock Australia Telescope National Facility, Sydney,
  (http://www.atnf.csiro.au/computing/software/miriad/)

\bibitem[\protect\citename{Sault \& Wieringa }{1994}]{sw94}
Sault~R.~J., Wieringa~M.~H., 1994, A\&AS, 108, 585

\bibitem[\protect\citename{Sault }{1994}]{sau94}
Sault~R.~J., 1994, A\&AS, 108, 55

\bibitem[\protect\citename{Sault, Staveley-Smith \& Brouw }{1996}]{ssb96}
Sault~R.~J., Staveley-Smith~L., Brouw~W.~N., 1996, A\&AS, 120, 375

\bibitem[\protect\citename{Sault, Teuben \& Wright }{1995}]{stw95}
Sault~R.~J., Teuben~P.~J., Wright~M. C.~H., 1995, in Shaw~R., Payne~H.,
  Hayes~J., eds, Astronomical {D}ata {A}nalysis {S}oftware and {S}ystems {IV}.
\newblock ASP Conference Series, Volume 77, San Francisco, p.~433

\bibitem[\protect\citename{Seward \& Harnden~Jr. }{1982}]{sh82}
Seward~F.~D., Harnden~Jr.~F.~R., 1982, ApJ, 256, L45

\bibitem[\protect\citename{Seward {\rm et~al. }}{1983}]{shmc83}
Seward~F.~D., Harnden~Jr.~F.~R., Murdin~P., Clark~D.~H., 1983, ApJ, 267, 698

\bibitem[\protect\citename{Seward {\rm et~al. }}{1984}]{shss84}
Seward~F.~D., Harnden~Jr.~F.~R., Szymkowiak~A., Swank~J., 1984, ApJ, 281, 650

\bibitem[\protect\citename{Shaver \& Goss }{1970a}]{sg70c}
Shaver~P.~A., Goss~W.~M., 1970a, Aust.\,J.\,Phys.\,Astr.\,Supp., 14, 133

\bibitem[\protect\citename{Shaver \& Goss }{1970b}]{sg70b}
Shaver~P.~A., Goss~W.~M., 1970b, Aust.\,J.\,Phys.\,Astr.\,Supp., 14, 77

\bibitem[\protect\citename{Shaver {\rm et~al. }}{1982}]{sra+82}
Shaver~P.~A., Radhakrishnan~V., Anantharamaiah~K.~R., Retallack~D.~S.,
  Wamsteker~W., Danks~A.~C., 1982, AA, 106, 105

\bibitem[\protect\citename{Smith }{1988}]{smi88b}
Smith~A., 1988, in Roger~R.~S., Landecker~T.~L., eds, Supernova Remnants and
  the Interstellar Medium ({IAU} Colloquium 101).
\newblock Cambridge University Press, Cambridge, p.~119

\bibitem[\protect\citename{Srinivasan, Bhattacharya \& Dwarakanath
  }{1984}]{sbd84}
Srinivasan~G., Bhattacharya~D., Dwarakanath~K.~S., 1984, JA\&A, 5, 403

\bibitem[\protect\citename{Sulkanen \& Lovelace }{1990}]{sl90}
Sulkanen~M.~E., Lovelace~R. V.~E., 1990, ApJ, 350, 732

\bibitem[\protect\citename{Tamura {\rm et~al. }}{1996}]{tkyb96}
Tamura~K., Kawai~N., Yoshida~A., Brinkmann~W., 1996, Proc.\,Astr.\,Soc.\,Jap.,
  48, L33

\bibitem[\protect\citename{Taylor \& Cordes }{1993}]{tc93}
Taylor~J.~H., Cordes~J.~M., 1993, ApJ, 411, 674

\bibitem[\protect\citename{Trussoni {\rm et~al. }}{1996}]{tmc+96}
Trussoni~E., Massaglia~S., Caucino~S., Brinkmann~W., Aschenbach~B., 1996, AA,
  306, 581

\bibitem[\protect\citename{Turtle {\rm et~al. }}{1962}]{tpkp62}
Turtle~A.~J., Pugh~J.~F., Kenderdine~S., Pauliny-Toth~I. I.~K., 1962, MNRAS,
  124, 297

\bibitem[\protect\citename{Ulmer {\rm et~al. }}{1993}]{umw+93}
Ulmer~M.~P. {\rm et~al.}, 1993, ApJ, 417, 738

\bibitem[\protect\citename{van~den Bergh \& Kamper }{1984}]{vk84}
van~den Bergh~S., Kamper~K.~W., 1984, ApJ, 280, L51

\bibitem[\protect\citename{van Langevelde \& Cotton }{1990}]{vc90}
van Langevelde~H.~J., Cotton~W.~D., 1990, AA, 239, L5

\bibitem[\protect\citename{Whiteoak \& Green }{1996}]{wg96}
Whiteoak~J. B.~Z., Green~A.~J., 1996, A\&AS, 118, 329,
  (http://www.physics.usyd.edu.au/astrop/wg96cat)

\bibitem[\protect\citename{Wieringa {\rm et~al. }}{1993}]{wdj+93}
Wieringa~M.~H., de~Bruyn~A.~G., Jansen~D., Brouw~W.~N., Katgert~P., 1993, AA,
  268, 215

\bibitem[\protect\citename{Wilson {\rm et~al. }}{1992}]{wff+92}
Wilson~R.~B., Finger~M.~H., Fishman~G.~J., Meegan~C.~A., Paciesas~W.~S., 1992.
\newblock IAU Circ. No. 5429

\bibitem[\protect\citename{Woosley, Eastman \& Schmidt }{1998}]{wes98}
Woosley~S.~E., Eastman~R.~G., Schmidt~B.~P., 1998, ApJ, submitted
  (astro-ph/9806299)

\bibitem[\protect\citename{Woosley, Langer \& Weaver }{1993}]{wlw93}
Woosley~S.~E., Langer~N., Weaver~T.~A., 1993, ApJ, 411, 823

\bibitem[\protect\citename{Woosley, Langer \& Weaver }{1995}]{wlw95}
Woosley~S.~E., Langer~N., Weaver~T.~A., 1995, ApJ, 448, 315

\end{thebibliography}
\label{lastpage}

\end{document}